\documentclass[11pt]{article}

\makeatletter\renewcommand{\section}{\@startsection
{section}{1}{\z@}{-3.5ex plus -1ex minus
    -.2ex}{2.3ex plus .2ex}{\bf }}

\makeatletter\renewcommand{\subsection}{\@startsection{subsection}{2}{\z@}{-3.25ex
plus -1ex minus
   -.2ex}{1.5ex plus .2ex}{\it }}
\makeatletter\renewcommand{\subsubsection}{\@startsection{subsubsection}{3}{-2.45ex}{-3.25ex
plus -1ex minus -.2ex}{1.5ex plus .2ex}{\it }}
\renewcommand{\thesection}{\arabic{section}}

\renewcommand{\theequation}{\thesection.\arabic{equation}}
\makeatletter \@addtoreset{equation}{section}

\usepackage{graphicx}
\usepackage{epsfig} 
\usepackage{hyperref}
\newcommand{\be}{\begin{equation}}
\newcommand{\ee}{\end{equation}}
\newcommand{\bea}{\begin{array}}
\newcommand{\ea}{\end{array}}
\newcommand{\beqa}{\begin{eqnarray}}
\newcommand{\eeqa}{\end{eqnarray}}
\newcommand{\nn}{\nonumber}
\renewenvironment{thebibliography}[1]
     {\baselineskip=16pt plus 2pt minus 1pt
      \section*{\large\refname
        \@mkboth{\MakeUppercase\refname}{\MakeUppercase\refname}}%
     \list{\@biblabel{\@arabic\c@enumiv}}%
           {\settowidth\labelwidth{\@biblabel{#1}}%
            \leftmargin\labelwidth
            \advance\leftmargin\labelsep
            \@openbib@code
            \usecounter{enumiv}%
            \let\p@enumiv\@empty
            \renewcommand\theenumiv{\@arabic\c@enumiv}}%
      \sloppy
      \clubpenalty4000
      \@clubpenalty \clubpenalty
      \widowpenalty4000%
      \sfcode`\.\@m}

\let\fn\footnote
\renewcommand{\footnote}[1]{\linespread{1.1}\fn{#1}\linespread{1.29}}

\usepackage{amsfonts}
\usepackage{amssymb}
\usepackage{mathrsfs}
\usepackage{amsmath,amssymb}
\usepackage{bm}
\usepackage{caption}
\usepackage{cite}

\newcommand{\appendices}{\section*{Appendices}\setcounter{section}{0}\setcounter{subsection}{0} \setcounter{equation}{0}
\renewcommand{\thesection}{\Alph{section}.}
\renewcommand{\theequation}{\thesection\arabic{equation}}}

\def\tyng(#1){\hbox{\tiny$\yng(#1)$}}

\begin{document}
\begin{titlepage}

\vskip 2 em

\begin{center}
\centerline{{\Large \bf Equivariant Fields in an $SU({\cal N})$ Gauge theory with new }}
\centerline{{\Large \bf 
Spontaneously Generated Fuzzy Extra Dimensions}}
\vskip 1em

\vskip 0.5em

\centerline{\large \bf S. K\"{u}rk\c{c}\"{u}o\v{g}lu and G. \"{U}nal}

\vskip 2em

\centerline{\sl  Middle East Technical University, Department of Physics,}
\centerline{\sl Dumlupinar Boulevard, 06800, \c{C}ankaya, Ankara, Turkey}

\vskip 2em

{\sl kseckin@metu.edu.tr\,,} {\sl ugonul@metu.edu.tr \,} 
\end{center}

\vskip 5 em

\begin{quote}
\begin{center}
{\bf Abstract}
\end{center}
{\small

We find new spontaneously generated fuzzy extra dimensions emerging from a certain deformation of $N=4$ supersymmetric Yang-Mills (SYM) theory with cubic soft supersymmetry breaking and mass deformation terms. First, we determine a particular four dimensional fuzzy vacuum that may be expressed in terms of a direct sum of product of two fuzzy spheres, and denote it in short as $S_F^{2\, Int}\times S_F^{2\, Int}$. The direct sum structure of the vacuum is clearly revealed by a suitable splitting of the scalar fields in the model in a manner that generalizes our approach in \cite{Seckinson}. Fluctuations around this vacuum have the structure of gauge fields over $S_F^{2\, Int}\times S_F^{2\, Int}$,  and this enables us to conjecture the spontaneous broken model as an effective $U(n)$ $(n < {\cal N})$ gauge theory on the product manifold $M^4 \times S_F^{2\, Int} \times S_F^{2\, Int}$. We support this interpretation by examining the $U(4)$ theory and determining all of the $SU(2)\times SU(2)$ equivariant fields in the model, characterizing its low energy degrees of freedom. Monopole sectors with winding numbers $(\pm 1,0),\,(0,\pm1),\,(\pm1,\pm 1)$ are accessed from $S_F^{2\, Int}\times S_F^{2\, Int}$ after suitable projections and subsequently equivariant fields in these sectors are obtained. We indicate how Abelian Higgs type models with vortex solutions emerge after dimensionally reducing over the fuzzy monopole sectors as well.  A family of fuzzy vacua is determined by giving a systematic treatment for the splitting of the scalar fields and it is made manifest that suitable projections of these vacuum solutions yield all higher winding number fuzzy monopole sectors. We observe that the vacuum configuration $S_F^{2\, Int}\times S_F^{2\, Int}$ identifies with the bosonic part of the product of two fuzzy superspheres with $OSP(2,2)\times OSP(2,2)$ supersymmetry and elaborate on this unexpected and intriguing feature.

}
 
\end{quote}

\end{titlepage}

\setcounter{footnote}{0}

\newpage

\section{Introduction}

$N=4$ Supersymmetric Yang Mills (SYM) theory in four dimensions with $SU({\cal N})$ gauge symmetry group appears to have special standing in bridging string theory to quantum field theory (QFT). As a QFT it has several appealing properties, among which its conformal invariance and UV finiteness, may be indicated at first glance. It is invariant under $S$-duality, interchanging the coupling constants $g_{YM}$ and $\frac{4 \pi}{g_{YM}}$ and it plays a central role in gauge/gravity duality as it is the most prominent example on the conformal field theory (CFT) side for AdS/CFT correspondence \cite{D'Hoker:2002aw, Ramallo:2013bua}. However, it is generally considered that this theory is not realistic as it has too much symmetry.

One possible route for accessing phenomenologically viable models from $N = 4$ SYM is to consider its deformations, which supplement the purely quartic potential of the scalar field sector of the theory with cubic soft supersymmetry breaking (SSB) and quadratic mass deformation terms in the scalar fields \cite{Dorey:1999sj,Dorey:2000fc,Auzzi:2008ep, Chatzistavrakidis:2009ix, Steinacker:2014lma, Steinacker:2014eua, Steinacker:2015mia}. The latter break some or all of the supersymmetry as well as some of the global $SU(4)_R$ symmetry. These theories as well as the closely related YM matrix models \cite{Steinacker:2014lma, Austing:2001ib, Ydri:2016dmy} possess fuzzy vacuum solutions, which are generically given as direct sums of fuzzy spheres ${\cal S}_F^2 (:= \oplus S_F^2)$ or that of products of fuzzy spheres ${\cal S}^2_F \times {\cal S}^2_F$ ($:= \oplus S_F^2 \times S_F^2$). $N=1^*$ models \cite{Dorey:1999sj,Dorey:2000fc,Auzzi:2008ep} is an example of models falling into this category with fuzzy sphere vacua ${\cal S}_F^2$, while the model given in \cite{Chatzistavrakidis:2009ix} serves as an example with ${\cal S}^2_F \times {\cal S}^2_F$ type vacuum. A broader perspective is gained by first noting that the $N = 4$ SYM may be obtained by dimensionally reducing the $N =1$ SYM in ten dimensions to four dimensions (see, for instance \cite{Erdmenger}), while the dimensional reduction of the latter to $0+1$ dimensions leads to the BFSS matrix model \cite{Banks:1996vh} description of M-theory on flat backgrounds and the fuzzy sphere vacua ${\cal S}_F^2$ also emerge from the massive deformations of the BFSS theory, so called the BMN matrix model, which is proposed to give a non-perturbative description of the M-theory on maximally supersymmetric pp-wave backgrounds \cite{Berenstein:2002jq, Dasgupta:2002hx}. Thus, evidently, a detailed study of the vacua and low energy structure of the aforementioned deformed models is quite central  to assess their potential value from a phenomenological point of view. With this being one of our intensions, in the present paper, we determine a family of vacuum solutions of the form ${\cal S}^2_F \times {\cal S}^2_F$, including all the monopole sectors, and investigate the low energy physics in these vacua for the model introduced in \cite{Chatzistavrakidis:2009ix}. In order to clearly state our further motivations and purpose for doing so, we would like to briefly discuss how such fuzzy vacua may be interpreted as extra dimensions of an effective gauge theory emerging from the deformed $N = 4$ SYM. 

In \cite{Aschieri:2006uw} it was shown that $SU(\cal N)$ YM theory in Minkowski space $M^4$, coupled to a triplet of scalar fields in the adjoint representation of the gauge group dynamically develop extra dimensions in the form of a fuzzy sphere $S_F^2$. To be more precise, the potential term in the Lagrangian of this model spontaneously breaks the gauge symmetry and the vacuum expectation values of the scalar fields form the fuzzy sphere $S_F^2$, while the fluctuations around this vacuum turn out to be the gauge fields over $S_F^2$. Thus, after symmetry breaking, an effective gauge theory on the manifold $M^4 {\times S_F^2} $ with a gauge group which is a subgroup of $SU(\cal N)$ is conjectured to emerge. Construction of the tower of Kaluza-Klein (KK) modes of the gauge fields and an inspection of its low lying modes supports this interpretation. Nevertheless, there is another complementary approach in developing the effective gauge theory interpretation and understanding the low energy limit in this and a range of other models, which one of us (S.K.) has recently been engaged in investigating \cite{Seckin-Derek, seckin-PRD2, seckin-PRD, Seckinson}. This, so called, equivariant parametrization technique entails imposing proper symmetry conditions on the fields of the model so that they transform covariantly under the action of the symmetry group of the extra dimensions up to gauge transformations of the emergent model. This simply transcribes, in the case of $U(2)$ gauge theory over ${\cal M} \times S_F^2$, as determining the gauge fields which remain covariant under rotations of $S_F^2$ up to $U(2)$ gauge transformations. Following this approach endows us with the explicit equivariant parametrizations of all the fields in the model and provides evidence for the interpretation of such models as effective gauge theories, since, subsequently, an effective low energy action may be obtained by integrating out (i.e. tracing over) the fuzzy extra dimensions and dimensionally reducing the theory. The latter leads to Abelian Higgs type models with new vortex solutions. It is necessary to note here that, what we have just described is essentially an adaptation and application of the coset space dimensional reduction (CSDR) techniques discussed in \cite{Forgacs, Zoupanos, Zoupanos-Review} (See also, \cite{Aschieri:2003vy} in this context). As this has been discussed throughly in our previous work, we refrain here to make a rehash and refer the reader to the references \cite{Seckin-Derek, seckin-PRD2, seckin-PRD, Seckinson}\footnote{The results obtained in the context of Aharony-Bergman-Jafferis-Maldacena (ABJM) models \cite{Aharony:2008ug, Nastase:2009ny}, appear to have similarities with those of ours in \cite{Seckin-Derek, seckin-PRD, Seckinson}. ABJM models are $N=6$ SUSY, $U({\cal N}) \times U({\cal N})$ Chern-Simons gauge theories at the level $(k, -k)$ that come together with scalar and spinor fields in the bifundamental and fundamental representation, respectively, of its $SU(4)$ $R$-symmetry group. In \cite{Gomis:2008vc,Terashima:2008sy}, a massive deformation of this model, which preserves the $N=6$ SUSY, but breaks the $R$-symmetry to $SU(2) \times SU(2) \times U(1)_A \times U(1)_B \times {\mathbb Z}_2$ was investigated. It turns out that, this deformed model has vacuum solutions which are fuzzy sphere(s) in the bifundamental formulation realized in terms of the Gomis-Rodriguez-Gomez- Van Raamsdonk-Verlinde (GRVV) matrices \cite{Gomis:2008vc}. In \cite{Mohammed:2012gi,Mohammed:2012rd}, a certain parametrization for the fields in the bosonic sector of this model has been suggested and it was shown to yield a low energy model in which four complex scalar fields interact with a sextic potential.}. For other related results on equivariant dimensional reduction \cite{Popov-Szabo, Lechtenfeld1, Popov, Popov2, Dolan-Szabo, Landi-Szabo,Lechtenfeld:2003cq} may be consulted. 

Aforementioned deformations of $N = 4$ SYM may be viewed to constitute a set of other examples in this context. In fact, \cite{Chatzistavrakidis:2009ix} focuses on a particular deformation of $N = 4$ SYM with both SSB and mass deformation terms, which completely breaks the supersymmetry and the $SU(4)$ R-symmetry down to a global $SU(2) \times SU(2)$.  This model has vacuum solutions of the form\footnote{In fact, to access all possible vacuum configurations as we do in the present work, one has to refrain from adding a constraint term to the Lagrangian as encountered in \cite{Chatzistavrakidis:2009ix}, which essentially forces to select the vacuum $S_F^2 \times S_F^2$, (see also \cite{Seckinson} in this context).} ${\cal S}^2_F \times {\cal S}^2_F$ and to our knowledge, it was the only example, until very recently\footnote{In a recent article \cite{Steinacker:2014lma}, new $4$- and $6$-dimensional fuzzy vacuum configurations in SSB deformed $N = 4$ SYM has been reported.}, in which a $4$-dimensional fuzzy vacuum emerges from deformed $N = 4$ SYM models. In \cite{Chatzistavrakidis:2009ix} it was shown that ${\cal S}^2_F \times {\cal S}^2_F$ type vacuum with background monopole fluxes leads to fermionic zero modes and mirror fermions are found to emerge in the low energy limit. In \cite{seckin-PRD}, one of us inspected the low energy structure of the effective gauge theory on $M^4 \times S_F^2 \times S_F^2$ with $U(4)$ gauge symmetry using the equivariant parametrization techniques and found, after tracing over the fuzzy extra dimension, abelian Higgs type models with three independent complex and several real scalar fields with new generalized vortex solutions. A complete treatment of the vacuum solutions of this model with background monopole fluxes and the low energy physics around such vacua is still missing in the literature, and this is intended as one of the aims of our present work. Here we extend the novel approach recently introduced by one of us in \cite{Seckinson}, which not only gives us access to all the fuzzy monopole sectors, but also reveals a whole family of fuzzy vacua with additional novel properties in the low energy structure. From a geometrical point of view, these vacua may be viewed as stacks of concentric fuzzy D-branes carrying magnetic monopole fluxes, although not all the string theoretic aspects \cite{Blumenhagen:2006ci} may be captured within the current framework, as already noted in \cite{Chatzistavrakidis:2009ix}. Thus, it is possible to view the equivariant gauge field modes that we obtain in section $3$ (see the ensuing paragraph for a brief description) as the modes of the gauge fields living on the world-volume of these D-branes, which may perhaps provide us with a good link to relate the effective gauge theory and the string theoretic perspectives. Also, we find it worthwhile to remark that our results apply just as well to the scalar sector of YM $6$-matrix models \cite{Steinacker:2014lma, Austing:2001ib, Ydri:2016dmy} whose global $SO(6)$ symmetry could be broken to $SU(2) \times SU(2)$ by SSB and/or mass deformation terms, making our work well connected to the ongoing research in such string related matrix models.

Having stated our motivations and purpose, we would like to briefly state how our work is organized and summarize our essential findings. In section 2, we determine a vacuum solution of the deformed $N = 4$ SYM model that may be expressed in terms of a particular direct sum of product of fuzzy spheres. For brevity we denote this solution as $S_F^{2\, Int}\times S_F^{2\, Int}$. The bosonic part of this model has six scalars $(\Phi_a^L \,, \Phi_a^R)$ ($a=1,2,3$) transforming under the adjoint representation of $SU(\cal N)$ and the $(1,0) \oplus (0,1)$ of $SU(2) \times SU(2)$. In the same vein to the technique introduced in \cite{Seckinson}, we show that the structure of $S_F^{2\, Int} \times S_F^{2\, Int}$ may be clearly revealed by splitting the scalar fields as $\Phi^L_a=\phi^L_a + \Gamma_a^L$, $\Phi^R_a=\phi^R_a+\Gamma_a^R$ where the constituents $(\Gamma_a^L , \Gamma_a^R)$ are defined by utilizing the four scalar fields $\Psi^L_\alpha$, $\Psi^R_\alpha$ ($\alpha = 1,2$) and their Hermitian conjugates, which are still in the adjoint of the $SU({\cal N})$, but transforming under the $(\frac{1}{2},0) \oplus (0, \frac{1}{2})$ of the global symmetry group. Certain bilinear composites of $\Psi^{L , R}_\alpha$ transforming in the $(1,0) \oplus (0,1)$ representation of $SU(2) \times SU(2)$ give the definition of $(\Gamma_a^L , \Gamma_a^R)$. In this section we also show that the fluctuations about this vacuum have the structure of gauge fields over $S_F^{2\, Int} \times S_F^{2\, Int}$ and enables us to conjecture that  the spontaneous broken model is an effective $U(n)$ $(n < {\cal N})$ gauge theory on the product manifold $M^4 \times S_F^{2\, Int}\times S_F^{2\, Int}$. In section 3, we support our conjecture by examining the $U(4)$ theory and determining all of the $SU(2) \times SU(2)$-equivariant fields in the model, which constitute the low energy degrees of freedom corroborating with the effective gauge theory interpretation. At this stage, from purely group theoretical analysis we encounter with the equivariant spinor modes over $S_F^{2\, Int} \times S_F^{2\, Int}$. We explicitly construct these modes by utilizing the four component multiplet in the representation $(\frac{1}{2},0) \oplus (0, \frac{1}{2})$ of the global symmetry group. Clearly, these spinorial modes do not constitute independent dynamical degrees of freedom in the $U(4)$ effective gauge theory, but it is readily conceived that their suitable bilinears shall yield the equivariant gauge field modes on $S_F^{2\, Int} \times S_F^{2\, Int}$.  We access the monopole sectors with winding numbers $(\pm 1,0),\,(0,\pm1),\,(\pm1,\pm 1)$ from $S_F^{2\, Int}\times S_F^{2\, Int}$ after suitable projections and obtain the equivariant fields in these sectors as a subset of those of the parent model. The latter characterizes the low energy modes of the theory and making contact with the results of \cite{seckin-PRD}, we show that tracing over the fuzzy monopole sectors is bound to yield two decoupled Abelian Higgs-type models, each with a $U(1)^3$ gauge symmetry and static multivortex solutions characterized by three winding numbers. In section 4, by examining the splitting of the fields $(\Phi_a^L \,, \Phi_a^R)$ with the composite part involving a $k_1+k_2$ component multiplets transforming under the representation $(\frac{k_1-1}{2},0)\oplus (0,\frac{k_2-1}{2})$ of the global symmetry, we determine a family of fuzzy vacuum solutions. It is manifestly seen from our results that suitable projections of these vacuum solutions yield all higher winding number monopole sectors.

An unexpected feature of the vacuum configuration $S_F^{2\, Int}\times S_F^{2\, Int}$ that we determine is that it identifies with the bosonic part of the product of two fuzzy superspheres with $OSP(2,2)\times OSP(2,2)$ supersymmetry. This is especially interesting and deserves special attention, as it is completely unintended. In section $5$ we present it by examining the decomposition of typical superspin IRRs of $OSP(2,2) \times OSP(2,2)$ under $SU(2)\times SU(2)$ IRR and how a particular typical IRR of this group matches with the $SU(2)\times SU(2)$ IRR content of $S_F^{2\, Int}\times S_F^{2\, Int}$. In addition, we also give a construction of the generators of $OSP(2,2)\times OSP(2,2)$ in its nine-dimensional fundamental atypical representation, by projecting a relevant set of $16 \times 16$ matrices, which appear in our model as building blocks in the construction of the matrix algebra of the composite fields. We feel that further research is necessary to uncover whether there is a deeper physical reason for the appearance of this structure or it is simply accidental.

Considerable amount of the details of the analysis of sections 3 and 4 are relegated to the appendices A and B. In appendix C,  we discuss another vacuum solution to the model, where $4 \times 4$ matrices are attempted to be used as building blocks to construct $\Gamma_a^L$ and $\Gamma_a^R$ instead of the $16 \times 16 $ matrices used in section $2$. Although the structure we encounter looks superficially similar to the one we obtained in section $2$, we find that there is in fact a crucial difference; namely that the objects whose bilinears are  $\Gamma_a^L$ and $\Gamma_a^R$,  do not transform as $(\frac{1}{2},0) \oplus (0, \frac{1}{2})$ representation of $SU(2) \times SU(2)$. Nevertheless, treating this model as one in its own right we examine it in some detail. In particular, we find that the effective $U(4)$ gauge theory contains no equivariant spinor field modes at all. This is indeed what we expect and it corroborates very well with the fact indicated above, since, reversing the line of reasoning, the absence of equivariant spinor field modes implies that the introduction of the composite fields $\Gamma_a^L$ and $\Gamma_a^R$ with the desired symmetry properties is not possible. If the latter was possible, it would have contradicted the absence of the equivariant spinor field modes and vice versa.

There are a number of recent interesting articles within this general setting that we do not want pass without mention 
\cite{Zoupanos-1, Grosse:2010zq, Shahin1, Shahin2, Orfanidis}. In \cite{Zoupanos-1}, for instance, an orbifold projection of $N=4$ SYM theory have been introduced and extra dimensions which are twisted fuzzy spheres consistent with this orbifolding were found to emerge due to the presence of SSB terms in the model. Authors of \cite{Zoupanos-1} have also discussed, what these results may possibly entail for the standard model as well as the minimal supersymmetric standard model (MSSM). A higher dimensional $SU({\cal N})$ Yang-Mills matrix model, similar in vein to the Ishibashi-Kawai-Kitazawa-Tsuchiya (IKKT) model \cite{Ishibashi:1996xs} for type IIB string theory, was studied in \cite{Grosse:2010zq}. After an analysis of the spontaneous symmetry breaking patterns mediated by the appearance of fuzzy spheres, it was shown that remaining gauge symmetry $SU(3)_c \times SU(2)_L \times U(1)_Q$ couples to all fields of the standard model and the resulting low energy model is an extension of the latter. Models involving matrix valued fields in the adjoint of $SU(\cal N)$ have been proposed for inflation models in \cite{Shahin1,Shahin2}.

\section{Gauge Theory over ${\cal M} \times S_F^{2 \, Int} \times S_F^{2\, Int}$}

\subsection{The Model and some Preliminaries}{\label{section2.1}}

We consider a deformed $N=4$ SYM theory with $SU(\cal N)$ gauge symmetry. This model has six anti-Hermitian scalar fields $\Phi_i$ $(i = 1\,, \cdots \,, 6)$ transforming in the adjoint representation of $SU(\cal N)$:
\begin{equation}
\Phi_i \rightarrow U^{\dagger}\Phi_i U\,, \quad U \in SU(\cal N) \,.
\end{equation}
With the SSB and mass deformation terms the action in the bosonic sector is given as \cite{Chatzistavrakidis:2009ix, seckin-PRD} 
\begin{align}
S&=\int d^4 x \, Tr_{\cal N}\bigg (- \frac{1}{4g^2} F^{\dagger}_{\mu \nu} F^{\mu\nu} - (D_{\mu}\Phi_i)^\dagger (D^{\mu}\Phi_i)\bigg) -\frac{1}{g_L^2}V(\Phi^L) - \frac{1}{g_R^2}V(\Phi^R) - \frac{1}{g_{LR}^2}V(\Phi^{L,R}) \,,
\label{action}
\end{align}
where $F_{\mu\nu}= \partial_\mu A_\nu -\partial_\nu A_\mu + \lbrack A_\mu \,, A_\nu \rbrack$ is the curvature of the $su(\cal N)$ valued anti-Hermitian gauge fields $A_\mu$, $D_{\mu}\Phi_i = \partial_{\mu}\Phi_i + \lbrack A_{\mu},\Phi_i \rbrack $ is the covariant derivative of $\Phi_a$, $Tr_{\cal N} = {\cal N}^{-1} Tr$ is the normalized trace and 
\begin{equation}
\begin{split}
V(\Phi^L)&=Tr_{\cal N}F^{L\dagger}_{ab}F^L_{ab}\,,\quad F^L_{ab}=[\Phi^L_a,\Phi^L_b]-\epsilon_{abc}\Phi^L_c\,,
 \\V(\Phi^R)&=Tr_{\cal N}F^{R\dagger}_{ab}F^R_{ab}\,,\quad F^R_{ab}=[\Phi^R_a,\Phi^R_b]-\epsilon_{abc}\Phi^R_c\,,
\\V(\Phi^{L,R})&=Tr_{\cal N}F^{(L,R)\dagger}_{ab}F^{(L,R)}_{ab}\,,\quad F^{(L,R)}_{ab}=[\Phi^L_a,\Phi^R_b]\,,
\\\Phi^L_a&=\Phi_a\,,\quad \Phi^R_a=\Phi_{a+3}\,,\quad(a=1,2,3)\,.
\end{split}
\end{equation}  
Let us note that if replace the potential terms in (\ref{action}) with the purely quartic potential 
\begin{align}
V_{N=4}(\Phi)= - \frac{1}{4}g_{YM}^2\sum_{i,j}^6[\Phi_i,\Phi_j]^2 \,,
\end{align} 
then the action in (\ref{action}) reduces to the bosonic sector of the $N=4$ SYM, which possesses a global $SU(4)_R$ symmetry in addition to the local $SU(\cal N)$ \cite{D'Hoker:2002aw, Ramallo:2013bua}. 

The model in (\ref{action}) breaks the supersymmetry completely and the global $SU(4)_R$ down to a global $SU(2) \times SU(2)$. We observe that the scalar fields $\Phi_i \equiv (\Phi_a^L \,,\Phi_a^R)$ transform under the $(1,0) \oplus (0,1)$ representation of this global symmetry.

Following and generalizing the developments in \cite{Seckinson}, in this article, we are going to consider that $\Phi_a^L$ and $\Phi_a^R$ are split in the form
\begin{align}
&\Phi^L_a=\phi^L_a + \Gamma_a^L\,,\quad\Phi^R_a=\phi^R_a+\Gamma_a^R\,,
\label{structure}
\end{align}  
with the definitions
\begin{align}
\Gamma_a^L=-\frac{i}{2}\Psi^{L \dagger} \tilde{\tau}_a \Psi^L\,,\quad\Gamma_a^R=-\frac{i}{2}{\Psi}^{R \dagger} \tilde{\tau}_a {\Psi}^R\,, \quad \tilde{\tau}_a=\tau_a \otimes 1_{\cal N} \,, \quad \tau_a : \mbox{Pauli matrices}  \,,
\label{spinor}
\end{align}
where the scalar fields $\Psi^L$ and $\Psi^R$ are doublets of the global $SU(2)_L\times SU(2)_R$, transforming under its IRRs $(\frac{1}{2},0)$ and $(0,\frac{1}{2})$, respectively. Thus, we may form the $4$-component multiplet
\begin{align}
\Psi = 
\left (
\begin{array}{c}
\Psi^L \\
\Psi^R 
\end{array}
\right ) = 
\left (
\begin{array}{c}
\Psi_1^L \\
\Psi_2^L \\
\Psi_1^R \\
\Psi_2^R 
\end{array}
\right)
\,, 
\label{sp}
\end{align}
transforming under the representation $(\frac{1}{2},0) \oplus (0, \frac{1}{2})$ of the global symmetry group. We have that all the components $(\Psi^L_\alpha \,, \Psi^R_\alpha)$ ($\alpha=1,2$) of $\Psi$ are scalar fields; they are $\cal N \times \cal N$ matrices, transforming adjointly ($\Psi^{L,R}_\alpha \rightarrow U^\dagger \Psi^{L,R}_\alpha U$) under $SU({\cal N})$. Clearly, then $(\Gamma_a^L \,, \Gamma_a^R)$ are bilinears of $\Psi$'s transforming under the $(1,0) \oplus (0,1)$ of $SU(2)_L \times SU(2)_R$. Under the $SU({\cal N})$ gauge symmetry $(\Gamma_a^L \,, \Gamma_a^R)$ transform adjointly ($\Gamma^{L,R}_a \rightarrow U^\dagger \Gamma^{L,R}_a U$) as expected.

The doublets $\Psi^L$ and $\Psi^R$ have $4{\cal N}^2$ real degrees of freedom each, which in total appears to exceed the $6{\cal N}^2$ real degrees of freedom in $(\Phi_a^L \,, \Phi_a^R)$. Could this mean that there is something inconsistent about equations \eqref{structure} and \eqref{spinor}? The answer is no. To see why, let us recall that the adjoint action of $SU({\cal N})$ is composed of its left and the right actions. It is readily observed that, under the right action, $\Psi^L \rightarrow \Psi^L U$, $\Psi^R \rightarrow \Psi^R U$, we have $(\Gamma_a^L \,, \Gamma_a^R)$ transforming adjointly, while under the left action $\Psi^L \rightarrow U \Psi^L$, $\Psi^R \rightarrow V \Psi^R$, with $U,V \in SU({\cal N})$, we have $(\Gamma_a^L \,, \Gamma_a^R)$ remaining invariant. In other words, both $(\Psi^L \,, \Psi^R)$ and $(U \Psi^L \,, V \Psi^R)$ lead to the same $(\Gamma_a^L \,, \Gamma_a^R)$. Thus, what essentially enters into the definition of $(\Gamma_a^L \,, \Gamma_a^R)$ are the equivalence classes $(\Psi^L \,, \Psi^R) \sim (U \Psi^L \,, V \Psi^R)$. Since each of the unitary matrices $U \,, V \in SU({\cal N})$ have ${\cal N}^2$ real degrees of freedom, this means that each of $\Gamma_a^L$ and $\Gamma_a^R$ has $4{\cal N}^2 - {\cal N}^2 = 3{\cal N}^2$ real degrees of freedom, which yields exactly the same $6{\cal N}^2$ real degrees of freedom in $(\Gamma_a^L \,, \Gamma_a^R)$ as in $(\Phi_a^L \,, \Phi_a^R)$.

In fact, it can also be shown in a straightforward manner that the variations with respect to $\phi^{L,R}_a$ and $\Psi_\alpha^{L \, \dagger}$ and $\Psi_\alpha^{R \, \dagger}$ simply reproduce the same equations of motion as those that emerge from the variations\footnote{See Appendix A for details.} of $\Phi^{L,R}_a$\ indicating that no new degrees of freedom are introduced into the model by \eqref{structure}. This splitting is rather premature as it lacks any physical motivation at the present stage, but our reasons will become clear as we move forward and show that the model spontaneously develops fuzzy extra dimensions, which may be written as direct sums of the products $S_F^2 \times S_F^2$ as we shall now demonstrate.

\subsection{The Vacuum Configuration}{\label{section2.2}}
  
The potential terms in (\ref{action}) are positive definite, and therefore the minimum of the potential is given by the following equations
\begin{equation}
 F^L_{ab}=0\,,\quad F^R_{ab}=0\,,\quad F^{L,R}_{ab}=0\,.
 \label{minpot}
 \end{equation} 
 Solutions of these type of equations have been discussed in the literature \cite{Berenstein:2002jq, Chatzistavrakidis:2009ix, seckin-PRD}. In general, they are given by ${\cal N}\times {\cal N}$ matrices carrying reducible representations of $SU(2)\times SU(2)$ that decompose into direct sums of its IRRs. We want to consider such a solution to the equations (\ref{minpot}) in which we can take advantage of the splitting of the fields indicated in (\ref{structure}) and \eqref{spinor} in its construction. Let us emphasize that, the particular vacuum solution we want to construct this way exists regardless of our use of relations given in (\ref{structure}) and \eqref{spinor} as it is clear from our initial remark. Keeping these in mind, we can proceed to observe that the requirements in (\ref{structure}) and \eqref{spinor} naturally restrict the possible $SU(2)_L \times SU(2)_R$ representation that $(\Gamma_a^L,\,\Gamma_a^R)$ may carry to the one for which $(\Psi_\alpha^L,\Psi_\alpha^R)$ exists. In other words, $(\Gamma_a^L,\,\Gamma_a^R)$ may not be in some arbitrary representation of $SU(2)\times SU(2)$, since then the corresponding $(\Psi_\alpha^L,\Psi_\alpha^R)$ will not exist in general. Here we consider the only possible solution for which both $(\phi_a^L,\phi_a^R)$ and $(\Gamma_a^L,\,\Gamma_a^R)$ are nonzero matrices. 

We are going to show that the solution fulfilling the equations in (\ref{minpot}) with the structure given in (\ref{structure}) and (\ref{spinor}) may be written, assuming that $\cal N$ factors in the form ${\cal N}=(2\ell_L+1)\times(2\ell_R+1)\times 16\times n$, as 
\begin{equation}
   \label{vacuu}
   \begin{split}
    \Phi_a^L&=(X_a^{(2\ell_L+1)}\otimes \bm{1}^{(2\ell_R+1)}\otimes \bm{1}_{16}\otimes \bm{1}_n)+(\bm{1}^{(2\ell_L+1)}\otimes \bm{1}^{(2\ell_R+1)}\otimes {\Gamma_a^0}^L\otimes \bm{1}_n)\,,
  \\\Phi_a^R&=(\bm{1}^{(2\ell_L+1)}\otimes X_a^{(2\ell_R+1)}\otimes \bm{1}_{16}\otimes  \bm{1}_n)+(\bm{1}^{(2\ell_L+1)}\otimes \bm{1}^{(2\ell_R+1)}\otimes {\Gamma_a^0}^R\otimes \bm{1}_n )\,,
   \end{split}
  \end{equation}
up to gauge transformations $\Phi_i \rightarrow U^\dagger \Phi_i U$.
 
In (\ref{vacuu}), $(X_a^{(2\ell_L+1)},X_a^{(2\ell_R+1)})$ are the anti-Hermitian generators of $SU(2)_L\times SU(2)_R$ in the irreducible representation (IRR) $(\ell_L,\ell_R)$ and with the commutation relations
\begin{align}
[X_a^{(2\ell_L+1)},X_b^{(2\ell_L+1)}]=&\epsilon_{abc}X_c^{(2\ell_L+1)}\,,\quad [X_a^{(2\ell_R+1)},X_b^{(2\ell_R+1)}]=\epsilon_{abc}X_c^{(2\ell_R+1)}\,,\nonumber
\\&[X_a^{(2\ell_L+1)},X_b^{(2\ell_R+1)}]=0\,.
\label{fuzzycom}
\end{align}
$({\Gamma_a^0}^L,{\Gamma_a^0}^R)$ are conceived, for reasons that will become clear shortly, as $16\times 16$ anti-Hermitian matrices which satisfy the $SU(2)_L\times SU(2)_R$ commutation relations 
\begin{align}
&[{\Gamma_a^0}^L,{\Gamma_b^0}^L]=\epsilon_{abc}{\Gamma_c^0}^L\,,\quad [{\Gamma_a^0}^R,{\Gamma_b^0}^R]=\epsilon_{abc}{\Gamma_c^0}^R\,,\quad[{\Gamma_a^0}^L,{\Gamma_b^0}^R]=0 \,,
\label{SU(2)com}
\end{align} 
and form a reducible representation of $SU(2)_L\times SU(2)_R$.

We will now see that ${\Gamma_a^0}^L$ and ${\Gamma_a^0}^R$ can be written as bilinears of spinors carrying the IRR's $(\frac{1}{2},0)$ and $(0,\frac{1}{2})$, respectively. For this purpose, let us introduce four sets of fermionic annihilation-creation operators $(b_{\alpha} \,, b_{\alpha}^\dagger \,, c_{\alpha} \,, c_{\alpha}^\dagger)$ with the anticommutation relations
\be
\lbrace b_{\alpha},b_{\beta}^\dagger\rbrace=\delta_{\alpha \beta}\,, \quad \lbrace c_{\alpha},c_{\beta}^\dagger\rbrace=\delta_{\alpha \beta}\,, 
\ee
and all other anticommutators vanishing. They span the sixteen-dimensional Hilbert space $\cal H$ with the basis vectors
\begin{align}
&|n_1,\,n_2,\,n_3,\,n_4\rangle \equiv (b_1^\dagger)^{n_1}(b_2^\dagger)^{n_2}(c_1^\dagger)^{n_3}(c_2^\dagger)^{n_4}|0,\,0,\,0,\,0\rangle,
\end{align} with $n_1,n_2,n_3,n_4=0,1$. 

We can now take 
\begin{align}
{\Gamma_a^0}^L=-\frac{i}{2}\psi^{L \dagger} \tau_a \psi^L \,, \quad {\Gamma_a^0}^R=-\frac{i}{2}\psi^{R \dagger} \tau_a \psi^R\,,
\label{gam}
\end{align}
where 
\begin{align}
\psi^L := \left (
\begin{array}{c}
b_1 \\
b_2\\
\end{array}
\right )
\,,\quad 
\psi^R:=\left (
\begin{array}{c}
c_1\\
c_2
\end{array}
\right ) \,.
\label{Psichi}
\end{align}
It is easy to see that $({\Gamma_a^0}^L,{\Gamma_a^0}^R)$ fulfill the $SU(2)_L \times SU(2)_R$ commutation relations in (\ref{SU(2)com}). We furthermore have that 
\begin{equation}
\label{comspinor}
\begin{split}
 \lbrack \psi^L_{\alpha} \,, {\Gamma_a^0}^L \rbrack &=-\frac{i}{2}(\tau_a)_{\alpha \beta} \psi_{\beta}^L\,,\quad [{\psi_{\alpha}^{\dagger}}^L\,,{\Gamma_a^0}^L]=\frac{i}{2}(\tau_a)_{\beta \alpha}{\psi_{\beta}^{\dagger}}^L\,,\quad[\psi^L_{\alpha}\,,{\Gamma_a^0}^R]=0 \,,
 \\ [\psi^R_{\alpha},{\Gamma_a^0}^R]&=-\frac{i}{2}(\tau_a)_{\alpha \beta} \psi_{\beta}^R\,,\quad[{\psi_{\alpha}^{\dagger}}^R\,,{\Gamma_a^0}^R]=\frac{i}{2}(\tau_a)_{\beta \alpha}{\psi_{\beta}^{\dagger}}^R\,,\quad[\psi^R_{\alpha}\,,{\Gamma_a^0}^L]=0\,,
\end{split}
\end{equation}
therefore $\psi^L$ and $\psi^R$ carry the $(\frac{1}{2},0)$ and $(0,\frac{1}{2})$ IRRs of $SU(2)_L\times SU(2)_R$, respectively.
 
The quadratic Casimir of the representation spanned by $({\Gamma_a^0}^L,{\Gamma_a^0}^R)$ may be straightforwardly calculated to give 
  \begin{equation}
 C_2=({\Gamma_a^0}^L)^2+({\Gamma_a^0}^R)^2= \left (
\begin{array}{cccc}
\bm{0}_4 & 0 & 0  \\
0 & -\frac{3}{4}\bm{1}_8 & 0 \\
0 & 0 & -\frac{3}{2}\bm{1}_4 \\
\end{array}  
\right )
\label{casimi} \,,
\end{equation}
where we have used 
  \begin{align}
   ({\Gamma_a^0}^L)^2=-\frac{3}{4}N^L + \frac{3}{2}N_1^L N_2^L \,,\quad({\Gamma_a^0}^R)^2=-\frac{3}{4}N^R+\frac{3}{2}N_1^R N_2^R\,,
  \end{align}
with the number operators on the Hilbert space $\cal H$ given as 
\begin{equation}
\begin{split}
 &N_1^L=b_1^{\dagger}b_1\,,\quad N_2^L=b_2^{\dagger}b_2\,,\quad N^L=N_1^L+N_2^L\,, \\
&N_1^R=c_1^{\dagger}c_1\,,\quad N_2^R=c_2^{\dagger}c_2\,,\quad N^R = N_1^R + N_2^R \,,
\end{split}
\end{equation}
and we have taken the basis vectors of $\cal H$ oriented in the order $|0000\rangle\,,|0011\rangle\,,|0001\rangle\,,|0010\rangle$, \\
$|1100\rangle\,,  |1111\rangle\,,|1101\rangle\,,|1110\rangle\,,|0100\rangle\,,|0111\rangle\,,|0101\rangle\,,|0110\rangle\,, |1000\rangle\,,|1011\rangle\,,|1001\rangle\,,|1010\rangle$.

We infer from (\ref{casimi}) and the symmetry of (\ref{comspinor}) under the exchange of $L\leftrightarrow R$ that $({\Gamma_a^0}^L,{\Gamma_a^0}^R)$ has the IRR content expressed as the following direct sum of IRR's of $SU(2)_L\times SU(2)_R$: 
\begin{equation}
 \bm{4}(0,0)\oplus \bm{2}\big(\frac{1}{2},0\big)\oplus \bm{2}\big(0,\frac{1}{2}\big)\oplus\big(\frac{1}{2},\frac{1}{2}\big)\,.
 \label{modu}
\end{equation} 

It is also possible to express $({\Gamma_a^0}^L,{\Gamma_a^0}^R)$ as 
\begin{align}
 {\Gamma_a^0}^L=\Gamma_a^0\otimes \bm{1}_4\,,\quad{\Gamma_a^0}^R=\bm{1}_4\otimes \Gamma_a^0
 \label{gamm}\,,
\end{align}
where 
\begin{align}
 {\Gamma_a^0}&=-\frac{i}{2}\psi^\dagger\tau_a\psi\,,\quad
 \psi = \left (
\begin{array}{c}
\psi_1 \\
\psi_2
\end{array}
\right )
:=
\left (
\begin{array}{c}
d_1 \\
d_2
\end{array}
\right ) \,,
\end{align} 
where $d_{\alpha},\, d^{\dagger}_{\alpha},\,(\alpha=1,2)$ are fermionic annihilation and creation operators spanning the Hilbert space
${\cal H}_d$ with the basis vectors $| m_1 \,, m_2 \rangle = (d_1^\dagger)^{m_1}  (d_2^\dagger)^{m_2} | 0 \,, 0 \rangle$. We have $N=d_\alpha^\dagger d_\alpha$ and also that $[\Gamma_a^0 \,, N] = 0$. $\Gamma_a^0$ carries a reducible representation of $SU(2)$ which decomposes into IRRs of $SU(2)$ as $0_{\bm 0}\oplus 0_{\bm 2}\oplus \frac{1}{2}$ (where the subscripts $0$ and $2$ in $0_{\bm 0}$ and $0_{\bm 2}$ are the eigenvalues of the number operator $N$ for the $SU(2)$ singlets)\cite{Seckinson}. Since $\Gamma_a^0$ fulfill the $SU(2)$ commutation relations, it is clear that $({\Gamma_a^0}^L,{\Gamma_a^0}^R)$ as defined in (\ref{gamm}) fulfill the commutation relations in (\ref{SU(2)com}). It is easily observed that (\ref{gam}) and (\ref{gamm}) describe unitarily equivalent representations and (\ref{gamm}) indeed yields identically the same set of $({\Gamma_a^0}^L,{\Gamma_a^0}^R)$ as in Eq. (\ref{gam}) if the basis vectors of ${\cal H}_d$ are taken in the order $|0 \, ,0 \rangle \,,|1 \,,1 \rangle\,,|0 \,,1 \rangle \,, |1 \,,0 \rangle$.

Let us first give the two projectors 
\begin{equation}
P_0=\frac{{({\Gamma_a^0}})^2+\frac{3}{4}}{\frac{3}{4}}=1-N+2N_1N_2\,, \quad \quad P_{\frac{1}{2}}=-\frac{{({\Gamma_a^0}})^2}{\frac{3}{4}}=N-2N_1N_2\,,
\end{equation}
where $P_0$ projects to the singlets and $P_{\frac{1}{2}}$ projects to the doublet of ${{\Gamma_a^0}}$, and $N_1=d_1^{\dagger}d_1,\,N_2=d_2^{\dagger}d_2,\,N=N_1+N_2$. It is also possible to distinguish between the two inequivalent singlets, $0_{\bm 0}$ and $0_{\bm 2}$, using the projectors
\begin{equation}
\label{P}
\begin{split}
 &P_{0_{\bm 0}}=-\frac{1}{2}(N-2)P_0=1-N-N_1N_2\,,
 \\&P_{0_{\bm 2}}=\frac{1}{2}NP_0=N_1N_2=-\frac{1}{2}N+\frac{1}{2}P_{\frac{1}{2}}\,.
\end{split}
\end{equation}

We can now consider the $SU(2)_L\times SU(2)_R$ IRR representation content of (\ref{vacuu}). Clebsch-Gordan decomposition gives
\begin{align}
  &(\ell_L,\ell_R)\otimes \left(\bm{4}(0,0)\oplus\bm{2}\big(\frac{1}{2},0\big)\oplus \bm{2}\big(0,\frac{1}{2}\big)\oplus \big(\frac{1}{2},\frac{1}{2}\big)\right)\nonumber
  \\&\equiv \bm{4}(\ell_L,\ell_R)\oplus \bm{2}\big(\ell_L-\frac{1}{2},\ell_R\big)\oplus \bm{2}\big(\ell_L+\frac{1}{2},\ell_R\big)\oplus \bm{2}\big(\ell_L,\ell_R-\frac{1}{2}\big)\oplus \bm{2}\big(\ell_L,\ell_R+\frac{1}{2}\big)\nonumber
  \\&\oplus \big(\ell_L-\frac{1}{2},\ell_R-\frac{1}{2}\big)\oplus \big(\ell_L+\frac{1}{2},\ell_R-\frac{1}{2}\big) \oplus \big(\ell_L-\frac{1}{2},\ell_R+\frac{1}{2}\big)\oplus \big(\ell_L+\frac{1}{2},\ell_R+\frac{1}{2}\big).
  \label{vac2}
\end{align}
For convenience, we introduce the short-hand notation $D_a^L:=X_a^L+{\Gamma_a^0}^L\,,D_a^R:=X_a^R+{\Gamma_a^0}^R$ for the vacuum solutions (\ref{vacuu})

In accordance with the decomposition in (\ref{vac2}), a unitary transformation puts $( D_a^L, D_a^R)$ into the block diagonal form $({\cal D}_a^L,{\cal D}_a^R)\equiv (U^\dagger D_a^LU\,,U^\dagger D_a^RU)$ whose entries can be inferred from the casimir of IRR's appearing in (\ref{vac2}) and their multiplicities (see Appendix A)
Therefore, we may interpret the vacuum configuration of the gauge theory (\ref{action}) in terms of direct sums of $S_F^{2}\times S_F^{2}$ given as
\begin{align}
 S_F^{2 \, Int}\times S_F^{2\, Int}&:\equiv \bm{4}\left(S_F^2(\ell_L)\times S_F^2(\ell_R)\right)\oplus \bm{2}\left(S_F^2(\ell_L-\frac{1}{2})\times S_F^2(\ell_R)\right)\nonumber
 \\&\oplus \bm{2}\left(S_F^2(\ell_L+\frac{1}{2})\times S_F^2(\ell_R)\right)\oplus \bm{2}\left(S_F^2(\ell_L)\times S_F^2(\ell_R-\frac{1}{2})\right)\nonumber
 \\&\oplus \bm{2}\left(S_F^2(\ell_L)\times S_F^2(\ell_R+\frac{1}{2})\right)\oplus \left(S_F^2(\ell_L-\frac{1}{2})\times S_F^2(\ell_R-\frac{1}{2})\right)\nonumber
 \\&\oplus \left(S_F^2(\ell_L+\frac{1}{2})\times S_F^2(\ell_R-\frac{1}{2})\right)\oplus \left(S_F^2(\ell_L-\frac{1}{2})\times S_F^2(\ell_R+\frac{1}{2})\right)\nonumber
 \\&\oplus \left(S_F^2(\ell_L+\frac{1}{2})\times S_F^2(\ell_R+\frac{1}{2})\right)
\label{summand}\,.
\end{align}

Alluding to our initial remarks after equation \eqref{minpot}, it is necessary to stress once again that the vacuum solution given in \eqref{summand} exists, independent of the steps taken to construct it in this section, although it appears to be rather cumbersome to predict it without the given considerations. Conversely, we can state that the existence of the vacuum solution \eqref{summand}, may be used to motivate the splitting of the fields as given in the equations \eqref{structure}, \eqref{spinor} and \eqref{gam}, \eqref{Psichi}. In fact, this argument only indicates that such $\Psi^L$ and $\Psi^R$ are available\footnote{We also note that, it is not always possible to introduce $\Psi^L$ and $\Psi^R$ carrying the required symmetry properties. In Appendix C we discuss a situation of this sort.}  to define $(\Gamma_a^L, \Gamma_a^R)$. However, there is, another important fact that ensues from our results in section $3.1$, which makes the introduction of $\Psi^L$ and $\Psi^R$ a natural as well as a necessary one, and we will see this shortly.

To each summand occurring in (\ref{summand}) there corresponds a projection given in the form
\begin{align}
  &\Pi_{\alpha \beta}=\prod_{\gamma \ne\alpha,\, \delta\ne\beta }\frac{-(X_a^L+{{\Gamma_a^0}^L})^2-(X_a^R+{{\Gamma_a^0}^R})^2-\lambda^L_{\gamma}(\lambda^L_{\gamma}+1)-\lambda^R_{\delta}(\lambda^R_{\delta}+1)}{\lambda^L_{\alpha}(\lambda^L_{\alpha}+1)+\lambda^R_{\beta}(\lambda^R_{\beta}+1)-\lambda^L_{\gamma}(\lambda^L_{\gamma}+1)-\lambda^R_{\delta}(\lambda^R_{\delta}+1)},
  \label{projector}
\end{align}
where $\alpha,\beta,\gamma,\,\delta=0,+,-$ and $\lambda^L_{\alpha}\,,\lambda^R_{\alpha}$ take on the values $\ell_L\,,\ell_L\pm\frac{1}{2}\,,\ell_R\,,\ell_R\pm\frac{1}{2}$ respectively. {\bf This} gives nine projectors. Note that $\Pi_{\alpha \beta}$ does not resolve the repeated summands in (\ref{summand}). For instance, $\Pi_{00}$ projects to the sector $\bm{4}\left(S_F^2(\ell_L)\times S_F^2(\ell_R)\right)$. We will see, how the projection to each repeated summand is accomplished as we proceed. 

It is important to note that these projectors may be expressed, after a unitary transformation, in terms of the products of the projectors $\Pi_{\alpha}^L$ and $\Pi_{\beta}^R$, which are given as 
\begin{equation}
\label{projector2}
\begin{split}
 &\Pi^L_{\alpha}=\prod_{\gamma\ne\alpha}\frac{-(X_a^L+{{\Gamma_a^0}^L})^2-\lambda^L_{\gamma}(\lambda^L_{\gamma}+1)}{\lambda^L_{\alpha}(\lambda^L_{\alpha}+1)-\lambda^L_{\gamma}(\lambda^L_{\gamma}+1)}\,,
 \\&\Pi^R_{\beta}=\prod_{\delta\ne\beta}\frac{-(X_a^R+{{\Gamma_a^0}^R})^2-\lambda^R_{\delta}(\lambda^R_{\delta}+1)}{\lambda^R_{\beta}(\lambda^R_{\beta}+1)-\lambda^R_{\delta}(\lambda^R_{\delta}+1)}\,.
\end{split}
\end{equation}
From  (\ref{projector2}), we may find that $\Pi_0^L,\,\,\Pi_0^R$, $\Pi_{\pm}^L,\,\,\Pi_{\pm}^R$ take the form
\begin{equation}
\begin{split}
 &\Pi^L_0= \bm{1}^{(2\ell_L+1)}\otimes \bm{1}^{(2\ell_R+1)}\otimes P_{0}\otimes \bm{1}_4\otimes \bm{1}_n \,, 
 \\&\Pi^R_0= \bm{1}^{(2\ell_L+1)}\otimes \bm{1}^{(2\ell_R+1)}\otimes \bm{1}_4\otimes P_{0}\otimes \bm{1}_n,
 \\&\Pi_{\pm}^L = \frac{1}{2} (\pm i {Q}^L_I + \Pi^L_{\frac{1}{2}}) \,, \quad\Pi_{\pm}^R = \frac{1}{2} (\pm i { Q}^R_I + \Pi^R_{\frac{1}{2}})\,,
\end{split}
\end{equation}
where \begin{align}
 {Q}^L_I=i\frac{X_a^L{\Gamma_a^0}^L-\frac{1}{4}\Pi^L_{\frac{1}{2}}}{\frac{1}{2}(\ell_L+\frac{1}{2})}\,,\quad { Q}^R_I=i\frac{X_a^R{\Gamma_a^0}^R-\frac{1}{4}\Pi^R_{\frac{1}{2}}}{\frac{1}{2}(\ell_R+\frac{1}{2})}\,,
\end{align} and $\Pi^L_{\frac{1}{2}}= \Pi^L_{+}+\Pi^L_{-},\,\,\Pi^R_{\frac{1}{2}}= \Pi^R_++\Pi^R_-.$ 

As $\Pi_{\alpha \beta}$ and $\Pi^L_{\alpha}\Pi^R_{\beta}$ project to the same subspaces, they are unitarily equivalent, $\Pi_{\alpha \beta}=U^\dagger\Pi^L_{\alpha}\Pi^R_{\beta}U$, for some unitary matrix $U$. Using the notation $\Pi_{\alpha \beta}\equiv \Pi^L_{\alpha}\Pi^R_{\beta}$ to denote this equivalence, we can list these nine projections onto the distinct IRRs in (\ref{vac2}) as given in the table 1 below
\begin{table}\centering
    \begin{tabular}{c | c }
    Projector & To the Representation \\ \hline
    \\
    $\Pi_{00}\equiv \Pi_0^L\Pi_0^R$ & $\bm{4}(\ell_L,\ell_R)$ \\
    $\Pi_{0\pm}\equiv \Pi_0^L\Pi_{\pm}^R$ & $\bm {2}(\ell_L,\ell_R\pm\frac{1}{2})$ \\
    $\Pi_{\pm 0}\equiv\Pi_{\pm}^L\Pi_{0}^R$ & $\bm{2}(\ell_L\pm \frac{1}{2},\ell_R)$ \\
    $\Pi_{\pm \pm}\equiv\Pi_{\pm}^L\Pi_{\pm}^R$ & $(\ell_L\pm \frac{1}{2},\ell_R\pm \frac{1}{2})$ \\
    $\Pi_{\pm \mp}\equiv\Pi_{\pm}^L\Pi_{\mp}^R$ & $(\ell_L\pm \frac{1}{2},\ell_R\mp \frac{1}{2}) $ \\
    \label{proje}
   \end{tabular}
   \caption{Projections $\Pi_{\alpha \beta}$}
   \label{table1}
    \end{table}

It is possible to split $\Pi_0^L$ to the projectors $\Pi_{0_{\bm 0}}^L,\Pi_{0_{\bm 2}}^L$ and $\Pi_0^R$ to $\Pi_{0_{\bm 0}}^R,\Pi_{0_{\bm 2}}^R$, as
\begin{equation}
\begin{split}
 &\Pi_{0_{\bm 0}}^L=\bm{1}^{(2\ell_L+1)}\otimes \bm{1}^{(2\ell_R+1)}\otimes P_{0_{\bm 0}}\otimes \bm{1}_4\otimes \bm{1}_n\,,\quad \Pi_{0_{\bm 2}}^L=\bm{1}^{(2\ell_L+1)}\otimes \bm{1}^{(2\ell_R+1)}\otimes P_{0_{\bm 2}}\otimes \bm{1}_4\otimes \bm{1}_n\,,
 \\&\Pi_{0_{\bm 0}}^R=\bm{1}^{(2\ell_L+1)}\otimes \bm{1}^{(2\ell_R+1)}\otimes \bm{1}_4\otimes P_{0_{\bm 0}}\otimes \bm{1}_n\,,\quad \Pi_{0_{\bm 2}}^R=\bm{1}^{(2\ell_L+1)}\otimes \bm{1}^{(2\ell_R+1)}\otimes \bm{1}_4 \otimes P_{0_{\bm 2}}\otimes \bm{1}_n\,,
\end{split}
\end{equation} where $P_{0_{\bm 0}}\,,P_{0_{\bm 2}}$ are given in (\ref{P}). Taking the above splitting of $\Pi_0^L$ and $\Pi_0^R$ into account, we can resolve $\Pi_{00}\,,\Pi_{0\pm}\,,\Pi_{\pm 0}$ into the projections, which project onto subspaces carrying a single IRR as given in table 2
\begin{table}\centering
\label{prop}
    \begin{tabular}{c | c }
    Projector & To the Representation \\ \hline
    \\
    ${\Pi}_{0_{\bm 0}}^L{\Pi}_{0_{\bm 0}}^R$ & $(\ell_L,\ell_R)$ \\
    ${\Pi}_{0_{\bm 0}}^L{\Pi}_{0_{\bm 2}}^R$ & $(\ell_L,\ell_R)$ \\
    ${\Pi}_{0_{\bm 2}}^L{\Pi}_{0_{\bm 0}}^R$ & $(\ell_L,\ell_R)$ \\
    ${\Pi}_{0_{\bm 2}}^L{\Pi}_{0_{\bm 2}}^R$ & $(\ell_L,\ell_R)$ \\
    ${\Pi}_{0_{\bm 0}}^L{\Pi}_{\pm}^R$ & $(\ell_L,\ell_R\pm \frac{1}{2})$ \\
    ${\Pi}_{0_{\bm 2}}^L{\Pi}_{\pm}^R$ &  $(\ell_L,\ell_R\pm \frac{1}{2})$ \\
    ${\Pi}_{\pm}^L{\Pi}_{0_{\bm 0}}^R$ & $(\ell_L\pm \frac{1}{2},\ell_R)$ \\
    ${\Pi}_{\pm}^L{\Pi}_{0_{\bm 2}}^R$ & $(\ell_L\pm \frac{1}{2},\ell_R)$ \\
    $\Pi_{\pm}^L\Pi_{\pm}^R$ & $(\ell_L\pm \frac{1}{2},\ell_R\pm \frac{1}{2})$ \\
    $\Pi_{\pm}^L\Pi_{\mp}^R$ & $(\ell_L\pm \frac{1}{2},\ell_R\mp \frac{1}{2}) $ \\
  \end{tabular}
  \caption{Projections to all fuzzy subspaces in r.h.s. of (\ref{summand}).}
  \label{table2}
\end{table}
These constitute the $16$ projectors onto the fuzzy subspaces appearing in the right hand side of equation (\ref{summand}).

\subsection{Gauge Theory over $M^4 \times S_F^{2 \, Int} \times S_F^{2\, Int}$}

We may now turn our attention back to the vacuum configuration (\ref{vacuu}). The latter breaks the $SU({\cal N})$ gauge symmetry to a $U(n)$. Clearly, this is the commutant of $(\Phi_a^L,\Phi_a^R)$ given in (\ref{vacuu}).  In addition, the global symmetry is totally broken by the vacuum. However, we note that, it is still possible to combine a global rotation with a gauge transformation which leaves the vacuum invariant.

We may introduce the fluctuations $(A_a^L,A_a^R)$ about the vacuum as 
\begin{equation}
\label{fluctuations}
 \begin{split}
  &\Phi_a^L=X_a^L+{\Gamma_a^0}^L+A^L_a=D_a^L+A_a^L\,,
  \\& \Phi_a^R=X_a^R+{\Gamma_a^0}^R+A^R_a=D_a^R+A_a^R\,,
  \end{split}
 \end{equation}
 where $A_a^L,\,A_a^R\in u(2\ell_L+1)\otimes u(2\ell_R+1)\otimes u(4)\otimes u(4)\otimes u(n)$. 

 Evaluating $F_{ab}^L,\,F_{ab}^R,\,F_{ab}^{L,R}$, we find 
 \begin{equation}
 \label{fieldtensor}
\begin{split}
&F_{ab}^L = [D_a^L,A_b^L]-[D_b^L,A_a^L]+[A_a^L,A_b^L]-\epsilon_{abc}A_c^L \,,
\\&F_{ab}^R  = [D_a^R,A_b^R]-[D_b^R,A_a^R]+[A_a^R,A_b^R]-\epsilon_{abc}A_c^R\,,
\\&F_{ab}^{L,R} = [D_a^L,A_b^R]-[D_b^R,A_a^L]+[A_a^L,A_b^R]\,.
 \end{split}
 \end{equation}
This suggests that we can think of $A_a^L$ and $A_a^R$ as the six components of a $U(n)$ gauge field living on $S_F^{2\,Int}\times S_F^{2\,Int}$ including the two normal components. As it is well-known, in fuzzy gauge theories, it is not possible to completely eliminate the normal components of the gauge fields \cite{Karabali:2001te,Balachandran:2003ay, Book}. However, it is possible to impose gauge invariant conditions on the fields which eliminate these normal components in the commutative limit, $\ell_L,\, \ell_R\rightarrow \infty$. Following the approaches in \cite{Karabali:2001te,Balachandran:2003ay, Book}, we introduce the conditions
\begin{equation}
\label{constraint}
 \begin{split}
  &(X_a^L+{\Gamma_a^0}^L+A_a^L)^2=(X_a^L+{\Gamma_a^0}^L)^2=-(\ell_L+\gamma)(\ell_L+\gamma+1){\bm 1}_{(2(\ell_L+\gamma)+1)(4(2\ell_R+1)n)}\,,
  \\&(X_a^R+{\Gamma_a^0}^R+A_a^R)^2=(X_a^R+{\Gamma_a^0}^R)^2=-(\ell_R+\gamma)(\ell_R+\gamma+1){\bm 1}_{(2(\ell_R+\gamma)+1)(4(2\ell_L+1)n)}\,,
 \end{split}
 \end{equation}
where $\gamma=0,\pm \frac{1}{2}$. In the commutative limit, $\ell_L,\, \ell_R\rightarrow \infty$, (\ref{constraint}) yields the transversality condition on ${\Gamma_a^0}^L+A_a^L$ and ${\Gamma_a^0}^R+A_a^R$ to be
\begin{align}
 &\hat{x}_a^L({\Gamma_a^0}^L+A_a^L)\rightarrow -\gamma\,, \quad \hat{x}_a^R({\Gamma_a^0}^R+A_a^R)\rightarrow -\gamma\,,
\end{align} as long as $A_a^{L,R}$ are smooth and bounded for $\ell_L,\, \ell_R\rightarrow \infty$ and converge to $A_a^L(x)\,,A_a^R(x)$ in this limit. Here we have $i \frac{X_a^{L,R}}{\ell}\rightarrow \hat{x}^{L,R}_a$ as $\ell_L,\, \ell_R\rightarrow \infty$, with $(\hat{x}^{L}_a \,, \hat{x}^{R}_a)$ being the coordinates of $S^2\times S^2$.
  
To summarize, we have a $U(n)$ gauge theory on ${\cal M}\times S_F^{2 \, Int}\times S_F^{2\, Int}$. Writing  $A_M:=(A_{\mu},A_a)$, the field strength tensor takes the form $F_{MN}=(F_{\mu \nu}\,,F^L_{\mu a}\,,F^R_{\mu a}\,, F_{ab}^L\,,F_{ab}^R\,,F_{ab}^{L,R})$ with
\begin{equation}
\begin{split}
 &F_{\mu a}^L:=D_{\mu}\Phi_a^L=\partial_{\mu}A_a^L-[X_a^L+{\Gamma_a^0}^L,A_{\mu}]+[A_{\mu},A_a^L]\,,
 \\&F_{\mu a}^R:=D_{\mu}\Phi_a^R=\partial_{\mu}A_a^R-[X_a^R+{\Gamma_a^0}^R,A_{\mu}]+[A_{\mu},A_a^R]\,.
\end{split}
\end{equation} and the rest already given in after (\ref{action}) and in (\ref{fieldtensor}).

\section{The $SU(2)\times SU(2)$-equivariant $U(4)$ gauge theory}

\subsection{Symmetries and Construction of the Equivariant Fields} 

In this section, we investigate the $U(4)$ gauge theory on $M^4 \times S_F^{2 \, Int}\times S_F^{2\, Int}$. In order to construct the $SU(2)\times SU(2)$-equivariant gauge fields, we introduce $ SU(2)\times SU(2)\approx SO(4)$ symmetry generators under which $A_{\mu}$ is a scalar, $A_a^L,\,A_a^R$ are $SU(2)_L\,,SU(2)_R$ vectors and $\Psi_{\alpha}^L\,,\Psi_{\alpha}^R$ are $SU(2)_L\,,SU(2)_R$ spinors, respectively, up to $U(4)$ gauge transformations \cite{seckin-PRD}. Our anti-Hermitian symmetry generators are
\beqa
\begin{aligned}
 \omega_a^L=(X_a^{(2\ell_L+1)}\otimes \bm{1}^{(2\ell_R+1)}\otimes \bm{1}_{16}\otimes \bm{1}_4)+(\bm{1}^{(2\ell_L+1)}\otimes \bm{1}^{(2\ell_R+1)}\otimes {\Gamma_a^0}^L\otimes \bm{1}_4)
 \\-(\bm{1}^{(2\ell_L+1)}\otimes \bm{1}^{(2\ell_R+1)}\otimes \bm{1}_{16} \otimes i\frac{L_a^L}{2})\,,
 \\\omega_a^R=(\bm{1}^{(2\ell_L+1)}\otimes X_a^{(2\ell_R+1)}\otimes \bm{1}_{16}\otimes \bm{1}_4)+(\bm{1}^{(2\ell_L+1)}\otimes \bm{1}^{(2\ell_R+1)}\otimes {\Gamma_a^0}^R\otimes \bm{1}_4)
 \\-(\bm{1}^{(2\ell_L+1)}\otimes \bm{1}^{(2\ell_R+1)}\otimes \bm{1}_{16}\otimes i\frac{L_a^R}{2})\,.
 \end{aligned} 
 \eeqa 
and $\omega_a^L=X_a^L+{\Gamma_a^0}^L+\frac{i}{2}L_a^L\,,\omega_a^R=X_a^L+{\Gamma_a^0}^R+\frac{i}{2}L_a^R$ for short. $L_a^L$ and $L_a^R$ are chosen so that $\omega_a^L$ and $\omega_a^R$ satisfy 
\begin{align}
 &[\omega_a^L,\omega_b^L]=\epsilon_{abc}\omega_c^L\,,\quad [\omega_a^R,\omega_b^R]=\epsilon_{abc}\omega_c^R\,,\quad [\omega_a^L,\omega_b^R]=0\,.
 \label{omega}
\end{align}
$(L_a^L,L_a^R)$ carry the $(\frac{1}{2},\frac{1}{2})$ IRR of $SU(2)\times SU(2)$. These six anti-symmetric $SU(4)$ matrices generate a $SU(2)\times SU(2)$ subalgebra in $U(4)$. The remaining nine symmetric generators of $SU(4)$ may be taken as $L_a^L L_b^R$. Together with the identity matrix $\bm 1_4$, these $16$ matrices form a basis for the fundamental representation of $U(4)$. In a suitable basis, $L_a^L$ and $L_a^R$ may be written to satisfy \cite{seckin-PRD}
\begin{align}
& L_a^LL_b^L=i\epsilon_{abc}L_c^L+\delta_{ab}\bm{1}_4\,,\quad L_a^RL_b^R=i\epsilon_{abc}L_c^R+\delta_{ab}\bm{1}_4\,,
\end{align}so that they can be viewed as two sets of $4\times 4$ ``Pauli matrices''.

From these facts, it is readily seen that the symmetry generators $(\omega_a^L,\omega_a^R)$ have the $SU(2)\times SU(2)$ representation content
\begin{align}
 &(\ell_L,\ell_R)\otimes \left(\bm{4}(0,0)\oplus\bm{2}(\frac{1}{2},0)\oplus \bm{2}(0,\frac{1}{2})\oplus (\frac{1}{2},\frac{1}{2})\right)\otimes (\frac{1}{2},\frac{1}{2})\nonumber
 \\&\equiv \bm{4}[(\ell_L+\frac{1}{2},\ell_R+\frac{1}{2})\oplus (\ell_L+\frac{1}{2},\ell_R-\frac{1}{2})\oplus (\ell_L-\frac{1}{2},\ell_R+\frac{1}{2})\oplus (\ell_L-\frac{1}{2},\ell_R-\frac{1}{2})]\nonumber
 \\&\oplus \bm{2}[(\ell_L-1,\ell_R-\frac{1}{2})\oplus (\ell_L-1,\ell_R+\frac{1}{2})]\oplus \bm{4}[(\ell_L,\ell_R+\frac{1}{2})\oplus (\ell_L,\ell_R-\frac{1}{2})]\nonumber
 \\& \oplus\bm{2}[(\ell_L+1,\ell_R-\frac{1}{2})\oplus (\ell_L+1,\ell_R+\frac{1}{2})]\oplus \bm{2}[(\ell_L-\frac{1}{2},\ell_R-1)\oplus (\ell_L+\frac{1}{2},\ell_R-1)]\nonumber
 \\&\oplus \bm{4}[(\ell_L-\frac{1}{2},\ell_R)\oplus (\ell_L+\frac{1}{2},\ell_R)]\oplus \bm{2}[(\ell_L-\frac{1}{2},\ell_R+1)\oplus (\ell_L+\frac{1}{2},\ell_R+1)]\nonumber
 \\&\oplus (\ell_L-1,\ell_R-1)\oplus \bm{2}(\ell_L-1,\ell_R)\oplus \bm{2} (\ell_L,\ell_R-1)\oplus \bm{4} (\ell_L,\ell_R)\oplus (\ell_L+1,\ell_R-1)\nonumber
 \\&\oplus \bm{2}(\ell_L+1,\ell_R)\oplus (\ell_L-1,\ell_R+1)\oplus \bm{2}(\ell_L,\ell_R+1)\oplus (\ell_L+1,\ell_R+1)\,.
 \label{vacua}
\end{align}

 Adjoint action of $(\omega_a^L,\omega_a^R)$ implies the following $SU(2)\times SU(2)$-equivariance conditions
 \begin{equation}
 \label{conditions}
\begin{split}
 &[\omega_a^L,A_{\mu}]=0\,,\quad [\omega_a^L,A^L_b]=\epsilon_{abc}A^L_c\,,\quad [\omega_a^L,\Psi_{\alpha}^L]=\frac{i}{2}(\tau_a)_{\alpha \beta}\Psi_{\beta}^L\,,
 \\&[\omega_a^R,A_{\mu}]=0\,,\quad [\omega_a^R,A^R_b]=\epsilon_{abc}A^R_c\,,\quad [\omega_a^R,{\Psi}_{\alpha}^R]=\frac{i}{2}(\tau_a)_{\alpha \beta}{\Psi}_{\beta}^R\,,
 \\&[\omega_a^L,A_b^R]=0=[\omega_a^R, A_b^L]\,,\quad  [\omega_a^L, \Psi_{\alpha}^R]=0=[\omega_a^R, \Psi_{\alpha}^L]\,.
\end{split}
\end{equation}

For the $U(4)$ theory under investigation, we list the projectors and the subspaces to which they project in the table below
\begin{table}
    \begin{tabular}{c | c }\centering
    Projector & To the Representation \\ \hline
    \\
    ${\Pi}_{0_{\bm 0}}^L{\Pi}_{0_{\bm 0}}^R$ & $(\ell_L+\frac{1}{2},\ell_R+\frac{1}{2})\oplus (\ell_L+\frac{1}{2},\ell_R-\frac{1}{2})\oplus (\ell_L-\frac{1}{2},\ell_R+\frac{1}{2})\oplus (\ell_L-\frac{1}{2},\ell_R-\frac{1}{2})$ \\
    ${\Pi}_{0_{\bm 0}}^L{\Pi}_{0_{\bm 2}}^R$ & $(\ell_L+\frac{1}{2},\ell_R+\frac{1}{2})\oplus (\ell_L+\frac{1}{2},\ell_R-\frac{1}{2})\oplus (\ell_L-\frac{1}{2},\ell_R+\frac{1}{2})\oplus (\ell_L-\frac{1}{2},\ell_R-\frac{1}{2})$ \\
    ${\Pi}_{0_{\bm 2}}^L{\Pi}_{0_{\bm 0}}^R$ & $(\ell_L+\frac{1}{2},\ell_R+\frac{1}{2})\oplus (\ell_L+\frac{1}{2},\ell_R-\frac{1}{2})\oplus (\ell_L-\frac{1}{2},\ell_R+\frac{1}{2})\oplus (\ell_L-\frac{1}{2},\ell_R-\frac{1}{2})$ \\
    ${\Pi}_{0_{\bm 2}}^L{\Pi}_{0_{\bm 2}}^R$ & $(\ell_L+\frac{1}{2},\ell_R+\frac{1}{2})\oplus (\ell_L+\frac{1}{2},\ell_R-\frac{1}{2})\oplus (\ell_L-\frac{1}{2},\ell_R+\frac{1}{2})\oplus (\ell_L-\frac{1}{2},\ell_R-\frac{1}{2})$ \\
    ${\Pi}_{0_{\bm 0}}^L{\Pi}_{\pm}^R$ & $(\ell_L+\frac{1}{2},\ell_R\pm1)\oplus(\ell_L-\frac{1}{2},\ell_R\pm1)\oplus (\ell_L+\frac{1}{2},\ell_R)\oplus(\ell_L-\frac{1}{2},\ell_R) $ \\
    ${\Pi}_{0_{\bm 2}}^L{\Pi}_{\pm}^R$ &  $(\ell_L+\frac{1}{2},\ell_R\pm1)\oplus(\ell_L-\frac{1}{2},\ell_R\pm1)\oplus (\ell_L+\frac{1}{2},\ell_R)\oplus(\ell_L-\frac{1}{2},\ell_R) $ \\
    ${\Pi}_{\pm}^L{\Pi}_{0_{\bm 0}}^R$ & $(\ell_L\pm1,\ell_R+\frac{1}{2})\oplus (\ell_L\pm1,\ell_R-\frac{1}{2})\oplus (\ell_L,\ell_R+\frac{1}{2})\oplus (\ell_L,\ell_R-\frac{1}{2})$ \\
    ${\Pi}_{\pm}^L{\Pi}_{0_{\bm 2}}^R$ & $(\ell_L\pm1,\ell_R+\frac{1}{2})\oplus (\ell_L\pm1,\ell_R-\frac{1}{2})\oplus (\ell_L,\ell_R+\frac{1}{2})\oplus (\ell_L,\ell_R-\frac{1}{2})$ \\
    ${\Pi}_{\pm}^L{\Pi}_{\pm}^R$ & $(\ell_L\pm 1,\ell_R\pm1)\oplus(\ell_L\pm1,\ell_R) \oplus(\ell_L,\ell_R\pm1)\oplus(\ell_L,\ell_R) $ \\
    ${\Pi}_{\pm}^L{\Pi}_{\mp}^R$ & $(\ell_L\pm 1,\ell_R)\oplus(\ell_L\pm1,\ell_R\mp1) \oplus(\ell_L,\ell_R)\oplus(\ell_L,\ell_R\mp1) $ \\
\end{tabular}
\caption{}
\end{table}
In order to avoid the possibility of any notational confusion, we note that the representation content of $(\omega_a^L,\omega_a^R)$ includes the tensor product with the IRR $(\frac{1}{2},\frac{1}{2})$ as seen in the l.h.s. of (\ref{vacua}) and $\Pi^L_{\alpha}\Pi^R_{\beta}$ project to the subspaces as listed in the table 3 above, while in the absence of the gauge symmetry generators $(L_a^L , L_a^R)$, $\Pi^L_{\alpha}\Pi^R_{\beta}$ project to the subspaces as listed in table \ref{table2}.

We can find the dimension of solution space for $A_{\mu},\,\,A^L_a,\,\,A^R_a$ and $\Psi_{\alpha}^L,\,\,\Psi_{\alpha}^R$ using the Clebsch-Gordan decomposition of the adjoint action of $(\omega_a^L,\omega_a^R)$. The relevant part of this decomposition gives
\begin{equation}
 \bm{196}(0,0) \oplus \bm{336}(\frac{1}{2},0)\oplus \bm{336}(0,\frac{1}{2})\oplus\bm{420}(1,0)\oplus\bm{420}(0,1) \cdots \,.
 \label{cgdeco}
\end{equation}
This means that there are $196$ equivariant scalars (i.e rotational invariants under $(\omega_a^L,\omega_a^R)$), $336$ equivariant spinors in each of the IRRs $(\frac{1}{2}\,,0)$ and  $(0\,,\frac{1}{2})$ and $420$ vectors in each of the IRRs $(1\,,0)$ and $(0\,,1)$. Employing the matrices 
\begin{gather}
S_i^L= {\bm 1}_{2 \ell_L+1} \otimes  {\bm 1}_{2 \ell_R+1} \otimes s_i \otimes {\bm 1}_4\otimes {\bm 1}_4\,,\quad  S_i^R= {\bm 1}_{2 \ell_L+1} \otimes  {\bm 1}_{2 \ell_R+1}  \otimes {\bm 1}_4 \otimes s_i \otimes {\bm 1}_4\,, \nn \\
s_i =
\left ( 
\begin{array}{cc}
\sigma_i & 0_2 \\
0_2 & 0_2
\end{array}
\right ) \,, \quad i = 1 \,, 2 \,,
\label{i1}
\end{gather}
\begin{equation}
\begin{split}
&Q^L_B=\frac{X^L_aL^L_a-\frac{i}{2}\bm 1}{\ell_L+\frac{1}{2}}\,,\quad Q_{0_{\bm 0}}^L={\Pi}_{0_{\bm 0}}^LQ_B^L\,,\quad Q_{0_{\bm 2}}^L={\Pi}_{0_{\bm 2}}^LQ_B^L\,,
\\&Q^L_+=\frac{1}{4\ell_L(\ell_L+1)}\Pi^L_+ \left( (2\ell_L+1)^2Q^L_B+i\right)\Pi^L_+\,,
\\&Q^L_-=\frac{1}{4\ell_L(\ell_L+1)}\Pi^L_- \left( (2\ell_L+1)^2Q^L_B-i\right)\Pi^L_-\,,
\\&Q^L_F={\Gamma_a^0}^L L^L_a-i\frac{1}{2}\Pi^L_{\frac{1}{2}}\,,\quad Q^L_H=-i\frac{\epsilon_{abc}X_a^L{\Gamma^0_b}^LL_c^L}{\sqrt{\ell_L(\ell_L+1)}}-\frac{1}{2}Q^L_{BI}+i\frac{1}{2}\Pi^L_{\frac{1}{2}}\,,
 \\&Q^L_{BI}=i\frac{(\ell_L+\frac{1}{2})^2\lbrace Q^L_B,Q^L_I\rbrace+\frac{1}{2}\Pi^L_{\frac{1}{2}}}{2\ell_L(\ell_L+1)}\,,\quad
Q_{S_i}^L = \frac{X_a^L S_i^L L_a^L - \frac{i}{2} S_i^L}{\ell_L + \frac{1}{2}}\,,
\end{split}
\label{i2}
\end{equation}
and $L\rightarrow R$ in (\ref{i2}) for the right constituents, a judicious choice of a basis for the equivariant scalars can be made so that they are ``idempotents'' in the subspace they live in, and they can be listed as 
\begin{equation}
\begin{split}
 &\Pi_i^L\Pi_i^R,\quad\Pi_i^LS_k^R,\quad\Pi_i^LQ_j^R,\quad\Pi_i^LQ_{S_k}^R,\quad Q_j^L\Pi_i^R,\quad Q_j^LS_k^R,\quad Q_j^LQ_j^R,\quad Q_j^LQ_{S_k}^R,
 \\&Q_{S_k}^L\Pi_i^R,\quad Q_{S_k}^LS_k^R,\quad Q_{S_k}^LQ_j^R,\quad Q_{S_k}^LQ_{S_k}^R,\quad S_k^L\Pi_i^R,\quad S_k^LS_k^R,\quad S_k^LQ_j^R,\quad S_k^LQ_{S_k}^R,
 \label{196}
\end{split}
\end{equation}
where $i$ runs over ${0_{\bm 0}},{0_{\bm 2}},+,-$, $j$ runs over ${0_{\bm 0}},{0_{\bm 2}},+,-,H,F$ and $k$ takes on the values $1,2$ and no sum over repeated indices is implied. Full lists of the equivariant spinors and vectors are not our immediate corcern in what follows and therefore they are relegated to Appendix A.

We note that that the index $\alpha$ ($\alpha = 1,2$) of $\Psi^L_\alpha$ and $\Psi^R_\alpha$ implying the transformation properties of these fields under the global symmetry $SU(2)_L\times SU(2)_R$, becomes, after symmetry breaking, the spinor index on $S_F^{2 \, Int} \times S_F^{2\, Int}$, just like the index $a$ ($a = 1,2,3$) of $(\Phi_a^{L} \,, \Phi_a^{R})$ becomes the vector index. We stress that the pure group theoretical result in equation \eqref{cgdeco} predicts the presence of equivariant spinor fields in the IRRs $(\frac{1}{2},0)$ and $(0,\frac{1}{2})$ of the symmetry group $SU(2)_L \times SU(2)_R$ of the fuzzy extra dimensions $S_F^{2 \, Int} \times S_F^{2\, Int}$. Their explicit construction, as listed in \eqref{sspi}, is only facilitated by the splittings of $\Phi^L$ an $\Phi^R$ in \eqref{structure} and \eqref{spinor}. As it should be already clear from our discussions in subsections \ref{section2.1} and \ref{section2.2}, these spinorial modes do not constitute independent dynamical degrees of freedom in the $U(4)$ effective gauge theory. Taking suitable bilinears of these spinors, we may construct all the equivariant gauge field modes on $S_F^{2\, Int} \times S_F^{2\, Int}$. In other words, it is in principle possible to express the "square roots" of the equivariant gauge field modes through these equivariant spinorial modes.

\subsection{Projection to the Monopole Sectors}

We gain much insight on the structure of the model by examining projections to its subsectors. We observe that $S_F{^{2\,Int}}\times S_F{^{2\,Int}}$ may be projected down to the monopole sectors
 \begin{align}
 &S_F^{2 \, \pm}\times S^{2}_F = \bigg (S_F^2(\ell_L)\times S_F^2(\ell_R)\bigg)\oplus \bigg( S_F^2(\ell_L\pm \frac{1}{2})\times S_F^2(\ell_R)\bigg )\label{case1}\,,
 \\\ &S_F^2\times S_F^{2 \, \pm} = \bigg (S_F^2(\ell_L)\times S_F^2(\ell_R)\bigg)\oplus \bigg ( S_F^2(\ell_L) \times S_F^2(\ell_R\pm \frac{1}{2})\bigg)\,,\label{case2}
 \\\  &S_F^{2 \, \pm}\times  S_F{^{2}}^{\pm}=\bigg (S_F^2(\ell_L)\times S_F^2(\ell_R)\bigg)\oplus\left (S_F^2(\ell_L\pm \frac{1}{2}) \times S_F^2(\ell_R\pm \frac{1}{2})\right)\label{case4}\,,
\end{align}
with the winding numbers $(\pm 1,0),\,(0,\pm 1),(\pm 1, \pm 1)$, respectively. This is indeed what we have been aiming at as indicated in the introduction. We can now probe the low energy structure of the $U(4)$ model in these monopole sectors by writing down their equivariant gauge field modes. In the next section , we will see how to systematically access all higher winding number monopole sectors.

Let us inspect each of the sectors briefly.

\vskip 1em

{\it i.} \underline{$S_F^{2 \, \pm}\times S^{2}_F$}: 

\vskip 1em

The projection (\ref{1pro}) to this sector is not unique, in the sense that there is in fact a set of projections which give the same monopole sector. We may consider, for instance, the projection
\begin{align}
\Pi_{0_{\bm 0}}^L\Pi_{0_{\bm 0}}^R+\Pi_{\pm}^L\Pi_{0_{\bm 0}}^R\,,
\label{1pro}
\end{align} 
We infer from (\ref{case1}) to which IRRs the projection (\ref{1pro}) restricts the direct sum given in the r.h.s. of (\ref{vac2}). After this projection, the number of equivariant fields are greatly reduced and they can be most easily found by working out the adjoint action of $(\omega_a^L,\omega_a^R)$, which in this subspace takes the simple form
\begin{equation}
\label{adjointrep}
\begin{split}
&\bigg[\left((\ell_L,\ell_R)\oplus (\ell_L\pm\frac{1}{2},\ell_R)\right)\otimes (\frac{1}{2},\frac{1}{2})\bigg]^{\otimes 2}\equiv \bm{8}(0,0)\oplus \bm{12}(\frac{1}{2},0) \oplus \bm{16}(1,0)\oplus \bm{16}(0,1)\cdots \,.
\end{split}
\end{equation} 
Thus, there are $8$ invariants which we read from (\ref{196}) as
\begin{align}
& \Pi_{0_{\bm 0}}^L\Pi_{0_{\bm 0}}^R\,,\quad \Pi_{\pm}^L\Pi_{0_{\bm 0}}^R\,,\quad \Pi_{0_{\bm 0}}^LQ_{0_{\bm 0}}^R\,,\quad \Pi_{\pm}^LQ_{0_{\bm 0}}^R\,,\quad Q_{0_{\bm 0}}^L\Pi_{0_{\bm 0}}^R\,,\quad Q_{\pm}^L\Pi_{0_{\bm 0}}^R\,,\quad Q_{0_{\bm 0}}^LQ_{0_{\bm 0}}^R\,,\quad  Q_{\pm}^LQ_{0_{\bm 0}}^R\,,
\label{inva}
\end{align}
$16$ vectors carrying the $(1,0)$ IRR 
\begin{equation}
\label{Lvec}
\begin{split}
 &\Pi_{0_{\bm 0}}^R[D_a^L,Q_{0_{\bm 0}}^L]\,,\quad \Pi_{0_{\bm 0}}^R Q_{0_{\bm 0}}^L[D_a^L,Q_{0_{\bm 0}}^L]\,,\quad \Pi_{0_{\bm 0}}^R \lbrace D_a^L,Q_{0_{\bm 0}}^L\rbrace\,,
  \\&Q_{0_{\bm 0}}^R[D_a^L, Q_{0_{\bm 0}}^L]\,,\quad Q_{0_{\bm 0}}^RQ_{0_{\bm 0}}^L[D_a^L,Q_{0_{\bm 0}}^L]\,,\quad Q_{0_{\bm 0}}^R\lbrace D_a^L, Q_{0_{\bm 0}}^L\rbrace\,,
  \\&\Pi_{0_{\bm 0}}^R[D_a^L, Q_{\pm}^L]\,,\quad \Pi_{0_{\bm 0}}^RQ_{\pm}^L[D_a^L,Q_{\pm}^L]\,,\quad \Pi_{0_{\bm 0}}^R \lbrace D_a^L,   Q_{\pm}^L\rbrace\,,
  \\&Q_{0_{\bm 0}}^R[D_a^L,Q_{\pm}^L]\,,\quad Q_{0_{\bm 0}}^RQ_{\pm}^L[D_a^L,Q_{\pm}^L]\,,\quad Q_{0_{\bm 0}}^R\lbrace D_a^L, Q_{\pm}^L\rbrace\,,
  \\&\Pi_{0_{\bm 0}}^R\Pi_{0_{\bm 0}}^L\omega_a^L\,,\quad \Pi_{0_{\bm 0}}^R\Pi_{\pm}^L\omega_a^L\,,\quad Q_{0_{\bm 0}}^R\Pi_{0_{\bm 0}}^L\omega_a^L\,,\quad Q_{0_{\bm 0}}^R\Pi_{\pm}^L\omega_a^L\,,
\end{split}
\end{equation}
and $16$ vectors in the $(0,1)$ IRR are 
\begin{equation}
\label{Rvec}
\begin{split}
 &\Pi_{0_{\bm 0}}^L[D_a^R,Q_{0_{\bm 0}}^R],\,\,\, \Pi_{0_{\bm 0}}^LQ_{0_{\bm 0}}^R[D_a^R,Q_{0_{\bm 0}}^R],\,\,\,\Pi_{0_{\bm 0}}^L \lbrace D_a^R, Q_{0_{\bm 0}}^R\rbrace,
  \\&Q_{0_{\bm 0}}^L[D_a^R, Q_{0_{\bm 0}}^R],\,\,\,Q_{0_{\bm 0}}^LQ_{0_{\bm 0}}^R[D_a^R,Q_{0_{\bm 0}}^R],\,\,\, Q_{0_{\bm 0}}^L\lbrace D_a^R, Q_{0_{\bm 0}}^R\rbrace,
  \\&\Pi_{\mp}^L[D_a^R, Q_{0_{\bm 0}}^R],\,\,\, \Pi_{\mp}^LQ_{0_{\bm 0}}^R[D_a^R,Q_{0_{\bm 0}}^R],\,\,\,\Pi_{\mp}^L \lbrace D_a^R, Q_{0_{\bm 0}}^R\rbrace,
  \\&Q_{\mp}^L[D_a^R,Q_{0_{\bm 0}}^R],\,\,\,Q_{\mp}^L Q_{0_{\bm 0}}^R[D_a^R,Q_{0_{\bm 0}}^R],\,\,\,  Q_{\mp}^L\lbrace D_a^R,Q_{0_{\bm 0}}^R\rbrace,
  \\&\Pi_{0_{\bm 0}}^L\Pi_{0_{\bm 0}}^R\omega_a^R,\,\,\, \Pi_{\mp}^L\Pi_{0_{\bm 0}}^R\omega_a^R,\,\,\,Q_{0_{\bm 0}}^L\Pi_{0_{\bm 0}}^R\omega_a^R,\,\,\,Q_{\mp}^L\Pi_{0_{\bm 0}}^R\omega_a^R\,.
 \end{split} 
\end{equation}
We see that there are $12$ equivariant spinor in the IRR $(\frac{1}{2},0)$
\begin{equation}
\label{spi}
\begin{split}
 &\Pi_{0_{\bm 0}}^R\Pi_{0_{\bm 0}}^L\beta_{\alpha}^LQ_{\pm}^L\,,\quad \Pi_{0_{\bm 0}}^RQ_{0_{\bm 0}}^L\beta_{\alpha}^L\Pi_{\pm}^L\,,\quad \Pi_{0_{\bm 0}}^RQ_{0_{\bm 0}}^L\beta_{\alpha}^LQ_{\pm}^L\,,\quad Q_{0_{\bm 0}}^R\Pi_{0_{\bm 0}}^L\beta_{\alpha}^LQ_{\pm}^L\,,\quad Q_{0_{\bm 0}}^RQ_{0_{\bm 0}}^L\beta_{\alpha}^L\Pi_{\pm}^L\,,
 \\&Q_{0_{\bm 0}}^RQ_{0_{\bm 0}}^L \beta_{\alpha}^LQ_{\pm}^L\,,\quad \Pi_{0_{\bm 0}}^R \Pi_{\pm}^L\beta_{\alpha}^L S_2^L\,,\quad \Pi_{0_{\bm 0}}^R \Pi_{\pm}^L\beta_{\alpha}^L Q_{s2}^L\,,\quad\Pi_{0_{\bm 0}}^R Q_{\pm}^L\beta_{\alpha}^L Q_{s2}^L\,,\quad Q_{0_{\bm 0}}^R \Pi_{\pm}^L\beta_{\alpha}^L S_2^L\,,
 \\&Q_{0_{\bm 0}}^R \Pi_{\pm}^L\beta_{\alpha}^L Q_{s2}^L\,,\quad Q_{0_{\bm 0}}^R Q_{\pm}^L\beta_{\alpha}^L Q_{s2}^L\,,
 \end{split}
\end{equation}
and due to the form of this monopole sector, we find no equivariant spinors in the IRR $(0,\frac{1}{2})$. 

One, rather trivial alternative to \eqref{1pro} is to change $\Pi_{0_{\bm 0}}^R$ with $\Pi_{0_{\bm 2}}^R$ in (\ref{1pro}), this simply amounts to taking $\Pi_{0_{\bm 0}}^R\rightarrow \Pi_{0_{\bm 2}}^R\,,Q_{0_{\bm 0}}^R\rightarrow Q_{0_{\bm 2}}^R$ in (\ref{inva}),(\ref{Lvec}),(\ref{Rvec}) and (\ref{spi}). Another choice is the projector
\begin{align}
  \Pi_{0_{\bm 0}}^L\Pi_{0_{\bm 0}}^R+\Pi_{\pm}^L\Pi_{0_{\bm 2}}^R\,.
  \label{2pro}
\end{align}
Equivariant fields in this case can be obtained in a similar fashion.

\vskip 1em

{\it i.} \underline{$S_F^{2} \times S^{2 \, \pm}_F$}: 

\vskip 1em

We observe that the only change in (\ref{adjointrep}) is the replacement of $(\frac{1}{2},0)$ with $(0,\frac{1}{2})$. Bearing this fact in mind, results in (\ref{inva}) to (\ref{2pro}) apply with the exchange $L\leftrightarrow R$.

\vskip 1em

{\it i.} \underline{$S_F^{2 \, \pm} \times S^{2 \, \pm}_F$}: 

\vskip 1em

To obtain this monopole sector we can use any one of the projections
 \begin{align}
  \Pi_{i}^L\Pi_{j}^R+\Pi_{\pm}^L\Pi_{\pm}^R\,,\quad i\,,j={0_{\bm 0}},{0_{\bm 2}}\,.
  \label{3pro}
 \end{align} 
In this case, the adjoint action of $(\omega_a^L,\omega_a^R)$ yields the representation content
\begin{align}
&\bigg[\left((\ell_L,\ell_R)\oplus (\ell_L\mp\frac{1}{2},\ell_R\mp \frac{1}{2})\right)\otimes (\frac{1}{2},\frac{1}{2})\bigg]^{\otimes 2}\equiv \bm{8}(0,0)\oplus \bm{16}(1,0)\oplus \bm{16}(0,1)\oplus \cdots\,.
\end{align}
We immediately observe that equivariant spinors are completely absent in this sector. Taking, for instance, $i\,,j={0_{\bm 0}}$ we find that $8$ scalars can be written as 
\begin{align}
&\Pi_{0_{\bm 0}}^L\Pi_{0_{\bm 0}}^R\,,\quad \Pi_{\pm}^L\Pi_{\pm}^R\,,\quad \Pi_{0_{\bm 0}}^LQ_{0_{\bm 0}}^R\,,\quad \Pi_{\pm}^LQ_{\pm}^R\,,\quad Q_{0_{\bm 0}}^L\Pi_{0_{\bm 0}}^R\,,\quad Q_{\pm}^L\Pi_{\pm}^R\,,\quad Q_{0_{\bm 0}}^LQ_{0_{\bm 0}}^R\,,\quad  Q_{\pm}^LQ_{\pm}^R\,,
\end{align} and $16$ vectors carrying the $(1,0)$ IRR are
\begin{align}
 &\Pi_{0_{\bm 0}}^R[D_a^L,Q_{0_{\bm 0}}^L]\,,\quad \Pi_{0_{\bm 0}}^R Q_{0_{\bm 0}}^L[D_a^L,Q_{0_{\bm 0}}^L]\,,\quad \Pi_{0_{\bm 0}}^R \lbrace D_a^L,Q_{0_{\bm 0}}^L\rbrace\,,\nonumber
  \\&Q_{0_{\bm 0}}^R[D_a^L, Q_{0_{\bm 0}}^L]\,,\quad Q_{0_{\bm 0}}^RQ_{0_{\bm 0}}^L[D_a^L,Q_{0_{\bm 0}}^L]\,,\quad  Q_{0_{\bm 0}}^R\lbrace D_a^L, Q_{0_{\bm 0}}^L\rbrace\,,\nonumber
  \\&\Pi_{\pm}^R[D_a^L, Q_{\pm}^L]\,,\quad  \Pi_{\pm}^RQ_{\pm}^L[D_a^L,Q_{\pm}^L]\,,\quad \Pi_{\pm}^R \lbrace D_a^L,Q_{\pm}^L\rbrace\,,\nonumber
  \\&Q_{\pm}^R[D_a^L,Q_{\pm}^L]\,,\quad Q_{\pm}^RQ_{\pm}^L[D_a^L,Q_{\pm}^L]\,,\quad Q_{\pm}^R\lbrace D_a^L, Q_{\pm}^L\rbrace\,,\nonumber
  \\&\Pi_{0_{\bm 0}}^R\Pi_{0_{\bm 0}}^L\omega_a^L\,,\quad \Pi_{\pm}^R\Pi_{\pm}^L\omega_a^L\,,\quad Q_{0_{\bm 0}}^R\Pi_{0_{\bm 0}}^L\omega_a^L\,,\quad Q_{\pm}^R\Pi_{\pm}^L\omega_a^L\,,
  \label{Lvec3}
\end{align} while the vectors carrying the $(0,1)$ representation follow from (\ref{Lvec3}) by the exchange $L\leftrightarrow R$.

\subsection{Parametrization of Fields and Comments on the Dimensional Reduced Action}

In all cases that we have discussed in this subsection, each summand of the projectors (given in (\ref{1pro}), (\ref{2pro}), (\ref{3pro}), etc.) splits the equivariant fields into mutually orthogonal subsectors under matrix product. For concreteness, let us briefly discuss the consequences of this fact for the sector given by the projection in (\ref{1pro}). We may write the parametrization of the fields $A_\mu$ as
\begin{multline}
 A_{\mu}=\frac{1}{2}a_{\mu}^1 \Pi_{0_{\bm 0}}^RQ_{0_{\bm 0}}^L +\frac{1}{2} a_{\mu}^2{\Pi}_{0_{\bm 0}}^L Q_{0_{\bm 0}}^R+\frac{i}{2}a_{\mu}^3{\Pi}_{0_{\bm 0}}^L\Pi_{0_{\bm 0}}^R+\frac{1}{2}ia_{\mu}^4Q_{0_{\bm 0}}^L  Q_{0_{\bm 0}}^R
 \\+\frac{1}{2}b_{\mu}^1 {\Pi}_{0_{\bm 0}}^R Q^L_{\pm}+\frac{1}{2}b_{\mu}^2\Pi^L_{\pm} Q_{0_{\bm 0}}^R+\frac{i}{2}b_{\mu}^3\Pi^L_{\pm}{\Pi}_{0_{\bm 0}}^R+\frac{1}{2}ib_{\mu}^4Q^L_{\pm} Q_{0_{\bm 0}}^R
 \end{multline}
 where $a_\mu^i$ and $b_\mu^i$, $(i=1,\cdots,4)$ are Abelian gauge fields
For $A_a^{L}$ we may write 
 \begin{align}
 A_a^L&=\frac{1}{2}(\chi_1+\chi_1^{\prime})\Pi_{0_{\bm 0}}^R[D_a^L,Q_{0_{\bm 0}}^L]+\frac{1}{2}(\chi_2+\chi_2^{\prime}-1)\Pi_{0_{\bm 0}}^RQ_{0_{\bm 0}}^L[D_a^L,Q_{0_{\bm 0}}^L]+i\frac{1}{4\ell_L}\chi_3 \Pi_{0_{\bm 0}}^R\lbrace D_a^L,Q_{0_{\bm 0}}^L\rbrace\nonumber
 \\&+\frac{1}{2\ell_L}\chi_4\Pi_{0_{\bm 0}}^L\Pi_{0_{\bm 0}}^R\omega_a^L+\frac{1}{2}(\chi_1-\chi_1^{\prime})iQ_{0_{\bm 0}}^R[D_a^L, Q_{0_{\bm 0}}^L]+\frac{1}{2}(\chi_2-\chi_2^{\prime})iQ_{0_{\bm 0}}^RQ_{0_{\bm 0}}^L[D_a^L,Q_{0_{\bm 0}}^L]\nonumber
 \\&+i\frac{1}{4\ell_L}\chi_3^{\prime}i Q_{0_{\bm 0}}^R\lbrace D_a^L, Q_{0_{\bm 0}}^L\rbrace+\frac{1}{2\ell_L}\chi_4^{\prime}iQ_{0_{\bm 0}}^R\Pi_{0_{\bm 0}}^L\omega_a^L\nonumber
 \\&+\frac{1}{2}(\varphi_1+\varphi_1^{\prime})\Pi_{0_{\bm 0}}^R[D_a^L, Q_{\pm}^L]+\frac{1}{2}(\varphi_2+\varphi_2^{\prime}-1)\Pi_{0_{\bm 0}}^RQ_{\pm}^L[D_a^L, Q_{\pm}^L]+i\frac{1}{4\ell_L}\varphi_3\Pi_{0_{\bm 0}}^R\lbrace D_a^L,Q_{\pm}^L\rbrace\nonumber
 \\&+\frac{1}{2\ell_L}\varphi_4 \Pi_{0_{\bm 0}}^R\Pi_{\pm}^L\omega_a^L+\frac{1}{2}(\varphi_1-\varphi_1^{\prime})iQ_{0_{\bm 0}}^R[D_a^L,Q_{\pm}^L]+\frac{1}{2}(\varphi_2-\varphi_2^{\prime})iQ_{0_{\bm 0}}^RQ_{\pm}^L[D_a^L,Q_{\pm}^L]\nonumber
 \\&+i\frac{1}{4\ell_L}\varphi_3^{\prime}iQ_{0_{\bm 0}}^R\lbrace D_a^L, Q_{\pm}^L\rbrace+\frac{1}{2\ell_L}\varphi_4^{\prime}iQ_{0_{\bm 0}}^R\Pi_{\pm}^L\omega_a^L\,,
 \label{Aleft}
 \end{align}
$(\chi_i, \chi_i^\prime, \varphi_i,\varphi_i^\prime)$, $(i=1,\cdots,4)$ are real scalar fields over $M^4$. Parametrization of $A_a^{R}$ may be written by taking $L \leftrightarrow R$ and replacing the scalars $(\chi_i, \chi_i^\prime, \varphi_i, \varphi_i^\prime)$ with the set $(\lambda_i, \lambda_i^\prime, \vartheta_i, \vartheta_i^\prime)$ in \eqref{Aleft}.
 
In $A_\mu$ first and second four terms are mutually orthogonal under matrix multiplication as they fall into two distinct projection sectors. Borrowing from the results of \cite{seckin-PRD}, we see that the low energy effective action of this model consists of two decoupled set of Abelian Higgs-type models with $U(1)^3$ gauge symmetry. In each subspace, we have three Abelian gauge fields coupled to four complex scalars which are $\chi=\chi_1+i\chi_2\,,\chi^\prime=\chi^\prime_1+i\chi^\prime_2$, $\lambda=\lambda_1+i\lambda_2\,,\lambda^\prime=\lambda^\prime_1+i\lambda^\prime_2$ in the first sector and $\varphi=\varphi_1+i\varphi_2\,,\varphi^\prime=\varphi^\prime_1+i\varphi^\prime_2$, $\vartheta=\vartheta_1+i\vartheta_2\,,\vartheta^\prime=\vartheta^\prime_1+i\vartheta^\prime_2$ in the second sector. Gauge fields $a_\mu^3$ and $b_\mu^3$ do not interact with any of the complex fields and they entirely decouple from the model in the $\ell_L\,,\ell_R\rightarrow \infty$ limit. Remaining eight real scalar fields in each sector interact only with the complex scalars in the respective sector they belong to. From the results of \cite{seckin-PRD} in the limit $\ell_L\,,\ell_R\rightarrow \infty$, the interaction potential in the first sector is given by sum of the three terms
\begin{align}
 V^L &= 4(|\chi|^2 +\frac{1}{4}(\chi_3 +\chi_3^\prime)-\frac{1}{4})^2 +4(|\chi^\prime|^2 +\frac{1}{4}(\chi_3 -\chi_3^\prime)-\frac{1}{4})^2  +2(\chi_3+\chi_3^\prime)^2 |\chi|^2 \nonumber \\
 &  \quad \quad + 2(\chi_3-\chi_3^\prime)^2 |\chi^\prime|^2 +\frac{1}{2}(\chi_4^2+{\chi_4^\prime}^2) \,,\nonumber
 \\V^R &= 4(|\lambda|^2 +\frac{1}{4}(\lambda_3 +\lambda_3^\prime)-\frac{1}{4})^2 +4(|\lambda^\prime|^2 +\frac{1}{4}(\lambda_3 -\lambda_3^\prime)-\frac{1}{4})^2 +2(\lambda_3+\lambda^\prime_3)^2 |\lambda|^2 \nonumber \\
 &  \quad \quad + 2(\lambda_3-\lambda_3^\prime)^2 |\lambda^\prime|^2 +\frac{1}{2}(\lambda_4+\lambda_4^\prime) \,,\nonumber
 \\V^{L,R}&=2(|\chi\lambda^\prime-\chi^\prime \lambda|^2+|\bar{\lambda}\chi-\chi^\prime \bar{\lambda}^\prime|^2)+\frac{1}{2}\big((|\chi|^2+|\chi^\prime |^2)({\lambda_3^\prime}^2 +{\lambda_4^\prime}^2) \nn \\
 & \quad \quad + (|\lambda|^2+|\lambda^\prime|^2)({\chi_3^\prime}^2+{\chi_4^\prime}^2)\big)\,,
 \label{pot1}
\end{align} while in the second sector, we have the potential given in the form (\ref{pot1}) with the substitutions $\chi_i\rightarrow \varphi_i\,,\chi_i^\prime\rightarrow \varphi_i^\prime$ and $\lambda_i\rightarrow \vartheta_i\,,\lambda_i^\prime\rightarrow \vartheta_i^\prime$.

Each sector possesses static multivortex solutions characterized by three winding numbers \cite{seckin-PRD}.

\section{Generalization of the Model with $k$-component Multiplets}

It is possible to search for other vacuum solutions for the action \eqref{action}. We may generalize the construction of section $2$ by replacing the doublets $\Psi^L$ and $\Psi^R$ in equation (\ref{spinor}) with $k_1$-, $k_2$-component multiplets of the global $SU(2)\times SU(2)$ as
\begin{align}
 \Psi^L= \left (
\begin{array}{c}
\Psi^L_1 \\
\Psi^L_2\\
\vdots \\
\Psi^L_{k_1}
\end{array}
\right )
\,,\quad 
\Psi^R=\left (
\begin{array}{c}
\Psi^R_1 \\
\Psi^R_2\\
\vdots \\
\Psi^R_{k_2}
\end{array}
\right )\,,\quad 
\Psi= \left (
\begin{array}{c}
\Psi^L\\
\Psi^R\\
\end{array}
\right) \,,
\end{align} 
transforming in its $(\frac{k_1-1}{2},0)$ and $(0,\frac{k_2-1}{2})$ IRR, respectively. Then, $\Psi$ is the $k_1+k_2$-component multiplet in the representation $(\frac{k_1-1}{2},0)\oplus (0,\frac{k_2-1}{2})$. Components $\Psi^L_\alpha\,,\Psi^R_\beta \in Mat({\cal N}) \,,(\alpha=1,\cdots \,, k_1)\,,(\beta=1\cdots \,, k_2)$ of $\Psi$ are scalar fields transforming in the adjoint representation of $SU(\cal N)$ as $\Psi^{L,R}_\alpha \rightarrow U^{\dagger}\Psi^{L,R}_\alpha U$. Bilinears $\Gamma_a^L$ and $\Gamma_a^R$ in $\Psi^L$ and $\Psi^R$ are defined similarly as before in the form
\begin{align}
 \Gamma_a^L=-\frac{i}{2}{\Psi^L}^\dagger \tilde{\lambda}^L_a \Psi^L\,,\quad \Gamma_a^R=-\frac{i}{2}{\Psi^R}^\dagger \tilde{\lambda}^R_a {\Psi^R}\,,\quad \tilde{\lambda}^{L,R}_a=\lambda^{L,R}_a\otimes \bm{1}_{\cal N}\,,
\end{align}
where now $\lambda_a^{L,R}$ are the generators of spin $(\frac{k_{L,R}-1}{2})$ representation of $SU(2)$.

In section $2$, we have seen that the vacuum configuration of our model can be written as the direct sum of products of fuzzy spheres whose structure is determined by the representation content of $({\Gamma_a^0}^L, {\Gamma_a^0}^R)$ with the corresponding doublet scalar fields taking the form given in (\ref{Psichi}). In order to generalize the latter, we need $k=k_1+k_2$ sets of annihilation-creation operators which satisfy
 \begin{align}
 &\lbrace b_{\alpha},b_{\beta}^\dagger\rbrace=\delta_{\alpha \beta}\,,\quad \alpha,\beta=1,\cdots, k_1\,,\quad  \lbrace c_{\rho},c_{\sigma}^\dagger\rbrace=\delta_{\rho \sigma}\,,\quad \rho,\sigma=1,\cdots ,k_2\,,
\end{align} 
with all other anticommutators vanishing. Thus, these operators span the $2^{k_1+k_2}$-dimensional Hilbert space  with the basis vectors
\begin{align}
 |n_1,\cdots, n_{k_1},m_1,\cdots, m_{k_2}\rangle={(b_1^{\dagger})}^{n_1}{(b_2^{\dagger})}^{n_2}\cdots {(b_{k_1}^{\dagger})}^{n_{k_1}}{(c_1^{\dagger})}^{m_1}{(c_2^{\dagger})}^{m_2}\cdots {(c_{k_2}^{\dagger})}^{m_{k_2}}|0,0\cdots,0\rangle\,,
\end{align} 
where $n_i,m_j=0,1,\,\, (i=1,\cdots, k_1,\,\,j=1,\cdots,k_2)$. For $\Psi^L=\psi^L$ and $\Psi^R=\psi^R$ with
\begin{align}
 \psi^L:= \left (
\begin{array}{c}
b_1 \\
\vdots\\
b_{k_1} \\
\end{array}
\right )
\,,\quad
\psi^R:=\left (
\begin{array}{c}
c_1 \\
\vdots\\
c_{k_2}
\end{array}
\right )\,,
\end{align} 
It is straightforward to show that ${\Gamma_a^0}^L=-\frac{i}{2}\psi^{L \dagger}\lambda_a\psi^L$ and ${\Gamma_a^0}^R=-\frac{i}{2}\psi^{R \dagger}\lambda_a\psi^R$ satisfy the $SU(2)\times SU(2)$ commutation relations and in addition fulfill
\begin{equation}
\begin{split}
 \lbrack \psi^L_{\alpha} \,, {\Gamma_a^0}^L \rbrack &=-\frac{i}{2}(\lambda_a)_{\alpha \beta} \psi_{\beta}^L\,,\quad[\psi^L_{\alpha}\,,{\Gamma_a^0}^R]=0\,,
 \\ [\psi^R_{\alpha},{\Gamma_a^0}^R]&=-\frac{i}{2}(\lambda_a)_{\alpha \beta} \psi_{\sigma}^R\,,\quad[\psi^R_{\alpha}\,,{\Gamma_a^0}^L]=0\,,
\end{split}
\end{equation} 
implying that $\psi_\alpha^L$ and $\psi_\alpha^R$ indeed carry the $(\frac{k_1-1}{2},0)$ and $(0,\frac{k_2-1}{2})$ IRRs, respectively.

In order to obtain the vacuum configuration in the present case, we have to first find out the $SU(2)\times SU(2)$ IRR content of $({\Gamma_a^0}^L, {\Gamma_a^0}^R)$. Number operators $N^L=b_{\alpha}^{\dagger}b_{\alpha}$ and $N^R=c_{\alpha}^{\dagger}c_{\alpha}$ commute with ${\Gamma_a^0}^L$ and ${\Gamma_a^0}^R$. This means that, the number of states in a given sector with eigenvalues $(n^L,n^R)$ $(n^L=(0,\cdots, k_1)\,,n^R=(0,\cdots, k_2))$ of $N^L$ and $N^R$ is equal to the dimension of one of the $SU(2) \times SU(2)$ IRR sectors occurring in the decomposition of the representation of $({\Gamma_a^0}^L, {\Gamma_a^0}^R)$ into the irreducibles of $SU(2)\times SU(2)$. Therefore, the IRRs of $SU(2)\times SU(2)$ that appear in $( {\Gamma_a^0}^L, {\Gamma_a^0}^R)$ may be labeled as
\begin{align}
(\ell^{k_1}_n,\ell^{k_2}_m)=\bigg(\frac{{k_1 \choose n}-1}{2} \,, \frac{{k_2 \choose m}-1}{2}\bigg) \,.
\label{multIRR}
\end{align} 
and the reducible representation carried by $({\Gamma_a^0}^L, {\Gamma_a^0}^R)$ decomposes into the direct sum
\begin{align}
L^{k_1\,k_2}:=\sum_{n=0}^{k_1}\sum_{m=0}^{k_2}\oplus (\ell^{k_1}_n,\ell^{k_2}_m)\,.
\label{summ}
\end{align} 
Since ${k_i \choose n} = {k_i \choose k-n}$, we see that $\ell_n^{k_i}=\ell_{k_i-n}^{k_i}$. As a consequence, not all the summands in (\ref{summ}) are distinct IRRs. Noting also that $\ell^{k_i}_{\frac{k_i}{2}}$ occurs only once for $k_i$ even, we may rewrite (\ref{summ}) as the direct sum of distinct IRRs together with its multiplicities as
\begin{align}
 L^{{k_1}_{even}\,{k_2}_{even}}&=(\ell_{\frac{k_1}{2}}^{k_1},\ell_{\frac{k_2}{2}}^{k_2})\oplus \bm{2}\sum_{n=0}^{\frac{k_1}{2}-1}\sum_{m=0}^{\frac{k_2}{2}}(\ell^{k_1}_n,\ell^{k_2}_m)\oplus \bm{2}\sum_{n=0}^{\frac{k_1}{2}}\sum_{m=0}^{\frac{k_2}{2}-1}(\ell^{k_1}_n,\ell^{k_2}_m)\,,\nonumber
 \\&=(\ell_{\frac{k_1}{2}}^{k_1},\ell_{\frac{k_2}{2}}^{k_2})\oplus \bm{4}\sum_{n=0}^{\frac{k_1}{2}-1}\sum_{m=0}^{\frac{k_2}{2}-1}(\ell^{k_1}_n,\ell^{k_2}_m)\oplus \bm{2}\sum_{n=0}^{\frac{k_1}{2}-1}(\ell^{k_1}_n,\ell^{k_2}_{\frac{k_2}{2}})\oplus \bm{2}\sum_{m=0}^{\frac{k_2}{2}-1}(\ell^{k_1}_{\frac{k_1}{2}},\ell^{k_2}_m)\,,\label{eveneven}
  \\ L^{{k_1}_{odd}\,{k_2}_{odd}}&= \bm{4}\sum_{n=0}^{\frac{k_1-1}{2}}\sum_{m=0}^{\frac{k_2-1}{2}}(\ell^{k_1}_n,\ell^{k_2}_m)\,,\label{oddodd}
  \\L^{{k_1}_{even}\,{k_2}_{odd}}&= \bm{4}\sum_{n=0}^{\frac{k_1}{2}-1}\sum_{m=0}^{\frac{k_2-1}{2}}(\ell^{k_1}_n,\ell^{k_2}_m)\oplus \bm{2}\sum_{m=0}^{\frac{k_2-1}{2}}(\ell^{k_1}_{\frac{k_1}{2}},\ell^{k_2}_m)\,.\label{evenodd}
 \end{align}
$ L^{{k_1}_{odd}\,{k_2}_{even}}$ can be obtained by taking $k_1\leftrightarrow k_2$ in equation (\ref{evenodd}).

With the assumption ${\cal N}=2^{k_1+k_2}(2\ell_L+1)(2\ell_R+1)n$, the vacuum configuration of our $SU(\cal N)$ gauge theory can be written as
 \begin{equation}
\begin{split}
  \Phi_a^L&=(X_a^{(2\ell_L+1)}\otimes \bm{1}^{(2\ell_R+1)}\otimes \bm{1}_{2^{k_1+k_2}}\otimes \bm{1}_n)+(\bm{1}^{(2\ell_L+1)}\otimes \bm{1}^{(2\ell_R+1)}\otimes {\Gamma_a^0}^L\otimes \bm{1}_n)
  \\\Phi_a^R&=(\bm{1}^{(2\ell_L+1)}\otimes X_a^{(2\ell_R+1)}\otimes \bm{1}_{2^{k_1+k_2}}\otimes  \bm{1}_n)+(\bm{1}^{(2\ell_L+1)}\otimes \bm{1}^{(2\ell_R+1)}\otimes {\Gamma_a^0}^R\otimes \bm{1}_n )\,,
 \end{split}
 \end{equation} 
up to $SU(\cal N)$ gauge transformations.

Clebsch-Gordan decomposition of the tensor products $(\ell_L,\ell_R)\otimes  L^{{k_1}_{even}\,{k_2}_{odd}}$, $ (\ell_L,\ell_R)\otimes L^{{k_1}_{even}\,{k_2}_{even}}$ and $(\ell_L,\ell_R)\otimes  L^{{k_1}_{odd}\,{k_2}_{odd}}$ reveal the vacuum configurations in terms of direct sums of $S_F^2\times S_F^2$. For instance, we have
\begin{align}
  {S_F^{2\,Int}}&_{k_1\,odd}  \times {S_F^{2\,Int}}_{k_2\,odd}:=\nonumber
  \\ \bm{4} & \sum_{n=0}^{\frac{k_1-1}{2}}\sum_{m=0}^{\frac{k_2-1}{2}}\bigg [S_F^2(\ell_L+\ell^{k_1}_n)\times S_F^2(\ell_R+\ell^{k_2}_m)\oplus \cdots \oplus S_F^2(\ell_L+\ell^{k_1}_n)\times S_F^2(|\ell_R-\ell^{k_2}_m|)\nonumber
 \\\oplus & S_F^2(\ell_L+\ell^{k_1}_n-1)\times S_F^2(\ell_R+\ell^{k_2}_m)\oplus \cdots \oplus S_F^2(\ell_L+\ell^{k_1}_n-1)\times S_F^2(|\ell_R-\ell^{k_2}_m|)\nonumber
 \\\oplus & \vdots \nonumber
 \\\oplus &S_F^2(|\ell_L-\ell^{k_1}_n|)\times S_F^2(\ell_R+\ell^{k_2}_m)\oplus \cdots \oplus S_F^2(|\ell_L-\ell^{k_1}_n|)\times S_F^2(|\ell_R-\ell^{k_2}_m|)\bigg]\,.
 \label{genodd}
\end{align} 
Remaining two cases are worked out explicitly in Appendix B.

We easily see from \eqref{genodd} and \eqref{geneven}, \eqref{genevenodd} that, all higher winding number monopole sectors may be obtained from suitable projections of  ${S_F^{2\,Int}}_{k_1} \times {S_F^{2\,Int}}_{k_2}$ in a systematic manner. As a quick example, let us consider the case with $k_1=k_2=3$. Then, $({\Gamma_a^0}^L,{\Gamma_a^0}^R)$ has the representation content
 \begin{align}
  \bm{4}[(0,0)\oplus (0,1)\oplus (1,0) \oplus (1,1)] \,,
 \end{align}
and the vacuum configuration takes the form
\begin{align}
{S_F^{2\,Int}}_{k_1=3}&\times {S_F^{2\,Int}}_{k_2=3}= {\bm 4} \bigg[{\bm 4} S_F^2(\ell_L)\times S_F^2(\ell_R) \oplus {\bm 2} S_F^2(\ell_L)\times S_F^2(\ell_R-1)\nonumber
\\\oplus &{\bm 2} S_F^2(\ell_L)\times S_F^2(\ell_R+1)\oplus {\bm 2} S_F^2(\ell_L-1)\times S_F^2(\ell_R)\oplus {\bm 2} S_F^2(\ell_L+1)\times S_F^2(\ell_R)\nonumber
\\\oplus& {\bm 2} S_F^2(\ell_L-1)\times S_F^2(\ell_R-1)\oplus {\bm 2} S_F^2(\ell_L-1)\times S_F^2(\ell_R+1)\nonumber
\\\oplus& {\bm 2} S_F^2(\ell_L+1)\times S_F^2(\ell_R-1)\oplus {\bm 2} S_F^2(\ell_L+1)\times S_F^2(\ell_R+1)\bigg]\,.
\end{align}
Monopole sectors with winding numbers $(0, \pm2), (\pm 2, 0), (\pm 2 \,, \pm 2), (\pm 2 \,, \mp 2)$ are all available through projections of ${S_F^{2\,Int}}_{k_1=3} \times {S_F^{2\,Int}}_{k_2=3}$. Sectors with winding numbers, such as $(n,n-1)$, appear through projections of ${S_F^{2\,Int}}_{k_1} \times {S_F^{2\,Int}}_{k_2}$ for $k_1 \neq k_2$.

Before closing this section, let us also remark that for the $U(4)$ gauge theory over ${S_F^{2\,Int}}_{k_1=3}\times {S_F^{2\,Int}}_{k_2=3}$ there are no equivariant spinors. This is quiet expected, since, for $k_1=k_2=3$, $\Psi^L$ and $\Psi^R$ transform under the IRRs $(1,0)$ and $(0,1)$ respectively  
and under the adjoint action of the symmetry generators we have 
\begin{align}
 [\omega_a^L, \Psi_b^L]=\frac{i}{2}{(\tilde{\lambda}_a)}_{bc}\Psi_c^L=\epsilon_{abc}\Psi_c^L\,,\quad  [\omega_a^R, \Psi_b^R]=\frac{i}{2}{(\tilde{\lambda}_a)}_{bc}\Psi_c^R=\epsilon_{abc}\Psi_c^R\,,
\end{align} 
since ${(\tilde{\lambda}_a)}_{bc}=-2i\epsilon_{abc}$ in the adjoint representation of $SU(2)$. Thus these equivariant field modes are one and the same as those obtained from the equivariance conditions on $\Phi^L_a$ and $\Phi^R_a$. From our results, we infer that the equivariant spinor fields over left and right fuzzy extra dimensions do exist only for both $k_1$ and $k_2$ even integers, while only left(right) spinor modes exist for $k_1$($k_2$) even only, and these modes do not exist at all for $k_1$ and $k_2$ both odd. 

\section{Relation between ${S_F^{2\,Int}} \times {S_F^{2\,Int}}$ and Fuzzy Superspace $S_F^{(2\,,2)}\times S_F^{(2\,,2)}$}

It is possible to identify the vacuum configuration given in equation (\ref{summand}) as the bosonic (even) part of the fuzzy space $S_F^{(2\,,2)}\times S_F^{(2\,,2)}$ with $OSP(2,2)\times OSP(2,2)$ symmetry. This observation makes the vacuum configuration $S_F^{(2\,,2)} \times S_F^{(2\,,2)}$ especially interesting since, it simply comes out naturally and we have in no way intended for it to emerge.

In order to reveal this relation, we have to write down the decomposition of IRRs of $OSP(2,2)\times OSP(2,2)$ under the $SU(2) \times SU(2)$ IRRs. 
Irreducible representations of $OSP(2,1)\times OSP(2,1)$ are characterized by two integer or half-integer numbers \newline
$({\cal J}_1,{\cal J}_2)_{OSP(2,1)\times OSP(2,1)}$ and it has the decomposition under the $SU(2)\times SU(2)$ IRRs as 

\begin{align}
 ({\cal J}_1,{\cal J}_2)=&\bigg[({\cal J}_1,{\cal J}_2)\oplus ({\cal J}_1-\frac{1}{2},{\cal J}_2)\oplus ({\cal J}_1,{\cal J}_2-\frac{1}{2})\oplus ({\cal J}_1-\frac{1}{2},{\cal J}_2-\frac{1}{2})\bigg]_{SU(2)\times SU(2)}\,.
 \label{Osp(2,1)}
\end{align}
Irreducible representations of $OSP(2,2)\times OSP(2,2)$ can be divided into two parts. These are the typical $({\cal J}_1,{\cal J}_2)_T$, and the atypical $({\cal J}_1,{\cal J}_2)_A$ representations. Typical representations $({\cal J}_1,{\cal J}_2)_T$ are reducible under the $OSP(2,1)\times OSP(2,1)$ IRRs as
\begin{align}
 ({\cal J}_1,{\cal J}_2)_T=&({\cal J}_1,{\cal J}_2)\oplus ({\cal J}_1-\frac{1}{2},{\cal J}_2)\oplus ({\cal J}_1,{\cal J}_2-\frac{1}{2})\oplus ({\cal J}_1-\frac{1}{2},{\cal J}_2-\frac{1}{2})\,,
 \label{Osp(2,2)}
\end{align} whereas the atypical ones are irreducible with respect to the $OSP(2,1)\times OSP(2,1)$ group and in fact $({\cal J}_1,{\cal J}_2)_A$ is equivalent to the IRR $({\cal J}_1,{\cal J}_2)$ of $OSP(2,1)\times OSP(2,1)$. All these facts follow from the generalization of the representation theory of $OSP(2,2)$ and $OSP(2,1)$, which is extensively discussed in \cite{Book, Balachandran:2002jf, Kurkcuoglu:2003ke}. With the help of equations (\ref{Osp(2,1)}) and (\ref{Osp(2,2)}), we see that $({\cal J}_1,{\cal J}_2)_T$ of $OSP(2,2)\times OSP(2,2)$ has the decomposition in terms of the IRRs of $SU(2)\times SU(2)$ as 
\begin{align}
 ({\cal J}_1,{\cal J}_2)_T=&\bigg[({\cal J}_1,{\cal J}_2)\oplus \bm{2}({\cal J}_1,{\cal J}_2-\frac{1}{2})\oplus \bm{2}({\cal J}_1-\frac{1}{2},{\cal J}_2)\oplus \bm{4}({\cal J}_1-\frac{1}{2},{\cal J}_2-\frac{1}{2})\nonumber
 \\&\oplus ({\cal J}_1-1,{\cal J}_2)\oplus \bm{2}({\cal J}_1-1,{\cal J}_2-\frac{1}{2})\oplus \bm{2}({\cal J}_1-\frac{1}{2},{\cal J}_2-1)\nonumber
 \\&\oplus ({\cal J}_1,{\cal J}_2-1)\oplus ({\cal J}_1-1,{\cal J}_2-1)\bigg]_{SU(2)\times SU(2)}\,,\quad \quad {\cal J}_1\,,{\cal J}_2\geq 1\,,
 \label{osp22}
\end{align}
while the representation $(\frac{1}{2},\frac{1}{2})_T$ decomposes as

\begin{align}
 (\frac{1}{2},\frac{1}{2})_T&\equiv(\frac{1}{2},\frac{1}{2})\oplus (0,\frac{1}{2})\oplus (\frac{1}{2},0)\oplus (0,0)\nonumber
 \\&\equiv \bigg[(\frac{1}{2},\frac{1}{2})+\bm{2}(0,\frac{1}{2})\oplus \bm{2}(\frac{1}{2},0)\oplus \bm{4} (0,0)\bigg]_{SU(2)\times SU(2)}\,.
 \label{coincide}
\end{align}
 
It is now easy to see that, for $({\cal J}_1,{\cal J}_2)_T\equiv (\ell_L+\frac{1}{2},\ell_R+\frac{1}{2})_T$, we obtain precisely the same IRR content from (\ref{osp22}) as the one that appears for the vacuum configuration given in (\ref{vac2}). This means that  $S_F^{2 \, Int}\times S_F^{2\, Int}$ can be identified with the bosonic part of the $OSP(2,2)\times OSP(2,2)$ fuzzy space $S_F^{(2\,,2)}\times S_F^{(2\,,2)}$ at the level $(\ell_L+\frac{1}{2},\ell_R+\frac{1}{2})_T$.

We further observe that $({\cal J}_1,{\cal J}_2) \equiv (\ell_L+\frac{1}{2},\ell_R+\frac{1}{2})$ IRR of $OSP(2,1)\times OSP(2,1)$ matches with a particular sector of the representation given in (\ref{vac2}) and allows us to identify 
\begin{multline}
\left(S_F^2(\ell_L+\frac{1}{2})\times S_F^2(\ell_R+\frac{1}{2})\right)\oplus\left(S_F^2(\ell_L)\times S_F^2(\ell_R+\frac{1}{2})\right)
\\\oplus \left(S_F^2(\ell_L+\frac{1}{2}\times S_F^2(\ell_R)\right)
\oplus \left(S_F^2(\ell_L)\times S_F^2(\ell_R)\right)\,,
\label{subsector}
\end{multline}
with the bosonic part of $OSP(2,1)\times OSP(2,1)$ fuzzy space $S_F^{(2\,,1)}\times S_F^{(2\,,1)}$. The subsector given in (\ref{subsector}) may be seen as the direct sum of two winding number $(1,0)$ monopole sectors as in (\ref{case1}) where one monopole sector differs from the other by the level of the right fuzzy spheres.

The superalgebra $osp(2,2)\times osp(2,2)$ has $16$ generators $\Lambda^i_M:=(\Lambda_a^i, \Lambda^i_\mu, \Lambda_8^i)\,,i=L\,,R$ which satisfy the graded commutation relations 
  \begin{equation}
\label{bosonic}
\begin{split} 
\lbrack \Lambda_a^i \,, \Lambda_b^i \rbrack &= i \varepsilon_{abc} \Lambda_c^i \,, \quad  \lbrack \Lambda_a^i \,, \Lambda_\mu^i \rbrack = \frac{1}{2} ({\Sigma_a})_{\nu \mu} \Lambda_\nu^i \,, \quad \lbrack \Lambda_a^i \,, \Lambda_8^i \rbrack = 0 \,, \quad
\\\lbrack \Lambda_8^i\,, \Lambda_\mu^i \rbrack &= \Xi_{\mu \nu} \Lambda_\nu^i \,,\quad \lbrace \Lambda_\mu^i \,, \Lambda_\nu^i \rbrace = \frac{1}{2} ({\cal C} \Sigma_a)_{\mu \nu} \Lambda_a^i + \frac{1}{4} (\Xi {\cal C})_{\mu \nu} \Lambda_8^i \,,
\end{split}
 \end{equation}
where 
\begin{align}
\Sigma_a = \left (
\begin{array}{cc}
\sigma_a & 0 \\
0& \sigma_a 
\end{array}
\right ) \,, \quad 
{\cal C } = \left (
\begin{array}{cc}
C & 0 \\
0& -C
\end{array}
\right ) \,, \quad 
\Xi =
\left (
\begin{array}{cc}
0 & I_2 \\
I_2 & 0 
\end{array}
\right ) \,, 
\end{align}
and $C$ is the two-dimensional Levi-Civita symbol and all the other graded commutation are zero. Reality condition implemented by the graded dagger operation on the generators reads
\begin{align}
\Lambda_a^\ddagger=\Lambda_a^\dagger=\Lambda_a\,,\quad  \Lambda_\mu^\ddagger = - {\cal C}_{\mu \nu} \Lambda_\nu \,, \quad \Lambda_8^\ddagger = \Lambda_8^\dagger = \Lambda_8 \,,
\end{align} 
for both the left and the right generators.

Using the representation theory of $osp(2,1)$ and $osp(2,2)$, it is rather straightforward to construct the nine-dimensional fundamental representation $(\frac{1}{2},\frac{1}{2})_A$ of  $osp(2,2)\times osp(2,2)$ which is at the same time the $(\frac{1}{2},\frac{1}{2})$ IRR of $osp(2,1)\times osp(2,1)$. Generators of the three-dimensional representation of $osp(2,2)$ may be written as
 \begin{align}
\lambda_a : =
\left ( 
\begin{array}{cc}
0 & 0 \\
0 & \frac{1}{2} \sigma_a
\end{array}
\right )  \,, \quad 
\lambda_4 : =  \frac{1}{2} \left (
\begin{array}{rrr}
0 & 0 & -1 \\
-1 &0 &0 \\
0 & 0 & 0
\end{array}
\right )
\,, \quad \lambda_5 : =  \frac{1}{2} \left (
\begin{array}{rrr}
0 & 1 & 0 \\
0 &0 & 0 \\
-1 & 0 & 0
\end{array}
\right ) \,,
\end{align}
\begin{align*}
\lambda_6 : =    \frac{1}{2} \left (
\begin{array}{rrr}
0 & 0 & -1 \\
1 &0 &0 \\
0 & 0 & 0
\end{array}
\right )
\,, \quad \lambda_7 : = \frac{1}{2} \left (
\begin{array}{rrr}
0 & 1 & 0 \\
0 &0 & 0 \\
1 & 0 & 0
\end{array}
\right ) \,,\quad
\lambda_8 : =  \left (
\begin{array}{rrr}
2 & 0 & 0 \\
0 & 1 & 0 \\
0 & 0 & 1
\end{array}
\right ) \,.
\end{align*}
Construction of these generators and a detailed exposition of the properties of the $osp(2,2)$ and $osp(2,1)$ superalgebras can be found in \cite{Book, Seckinson}. $16$ generators $(\Lambda^L_M \,, \Lambda^R_M)$ in the IRR $(\frac{1}{2},\frac{1}{2})_A$ can be given as\begin{align}
\Lambda^L_{M}\equiv \lambda_M\otimes \bm{1}_3\,,\quad \Lambda^R_a=\bm{1}_3\otimes \lambda_a\,,\quad \Lambda_{4\,,5}^R=\alpha\otimes \lambda_{4\,,5}\,,\quad \Lambda_{6\,,7}^R=-\alpha\otimes \lambda_{6\,,7}\,,\quad \Lambda_8^R=- {\bf1}_3\otimes \lambda_8\,,
\label{superfuzzy}
\end{align}
where $\alpha=3\bm{1}_3-2 \lambda_8$.

The matrices ${\Gamma_a^0}^L\,,{\Gamma_a^0}^R\,,b_\alpha\,,c_\alpha\,,b_\alpha^\dagger\,,c_\alpha^\dagger\,,N^L\,,N^R$ constitute a basis for the $16\times 16$ matrices acting on the sixteen-dimensional module corresponding to the representation space in (\ref{modu}), and coincides with that of (\ref{coincide}). We can make use of these matrices to construct generators of the representation $(\frac{1}{2},\frac{1}{2})_A$ given in (\ref{superfuzzy}). To do so, we should restrict to one of the nine-dimensional submodules with the representation content $(\frac{1}{2},\frac{1}{2})\oplus (0,\frac{1}{2})\oplus (\frac{1}{2},0)\oplus (0,0)$. Clearly, there exists a set of projectors which yield the same representation, and a particular projector from this set is
  \begin{equation}
  {\cal P}:={\cal P}_{0_{\bm 2}}^L {\cal P}_{0_{\bm 2}}^R+ {\cal P}_{0_{\bm 2}}^L{\cal P}_{\frac{1}{2}}^R+{\cal P}_{0_{\bm 2}}^R{\cal P}_{\frac{1}{2}}^L+{\cal P}_{\frac{1}{2}}^L{\cal P}_{\frac{1}{2}}^R\,,
 \end{equation}
where we have ${\cal P}_{0_{\bm 2}}^L=\bm{1}_4\otimes P_{0_{\bm 2}},\,{\cal P}_{\frac{1}{2}}^L= \bm{1}_4\otimes P_{\frac{1}{2}},\,\,{\cal P}_{0_{\bm 2}}^R=P_{0_{\bm 2}}\otimes \bm{1}_4,\,{\cal P}_{\frac{1}{2}}^R=P_{\frac{1}{2}}\otimes \bm{1}_4$. Using ${\cal P}$, we can restrict to the nine-dimensional submodule and subsequently get
\begin{equation}
\label{LambdaL}
\begin{split}
 \Lambda_1^L : &= - i  {\cal P}{\Gamma_1^0}^L\,,\quad \Lambda_2^L :=i  {\cal P}{\Gamma_2^0}^L\,,\quad \Lambda_3^L :=-i  {\cal P} {\Gamma_3^0}^L\,,\quad \Lambda_4^L : = - \frac{1}{2} ({\tilde b}_1 + {\tilde b}_2^\dagger)\,,
 \\\Lambda_5^L :& =  \frac{1}{2} ({\tilde b}^{\dagger}_1 - {\tilde b}_2)\,,\quad \Lambda_6^L : =  \frac{1}{2} ({\tilde b}_1 - {\tilde b}_2^{\dagger})\,,\quad \Lambda_7^L : =  \frac{1}{2} ({\tilde b}^{\dagger}_1 + {\tilde b}_2)\,,\quad \Lambda_8^L := {\cal P}N\,,
 \end{split}
\end{equation}
and
\begin{equation}
 \label{LambdaR}
\begin{split}
 \Lambda_1^R : &= - i  {\cal P}{\Gamma_1^0}^R\,,\quad \Lambda_2^R : =  i  {\cal P}{\Gamma_2^0}^R\,,\quad\Lambda_3^R : =  i  {\cal P}{\Gamma_3^0}^R\,,\quad \Lambda_4^R : = \frac{1}{2} ({\tilde c}_1 + {\tilde c}_2^\dagger)\,,
 \\\Lambda_5^R :& =  -\frac{1}{2} ({\tilde c}^{\dagger}_1 - {\tilde c}_2)\,,\quad \Lambda_6^R : =  \frac{1}{2} ({\tilde c}_1 - {\tilde c}_2^{\dagger})\,,\quad  \Lambda_7^R : = \frac{1}{2} ({\tilde c}^{\dagger}_1 + {\tilde c}_2)\,,\quad\Lambda_8^R : = -{\cal P}M\,,
\end{split} 
\end{equation}
where
\begin{align}
{\tilde b}_\alpha&={\cal P}b_\alpha {\cal P},\quad {\tilde b}_\alpha^\dagger={\cal P}b_\alpha^\dagger {\cal P},\quad{\tilde c}_\alpha={\cal P}c_\alpha {\cal P}, \quad {\tilde c}_\alpha^\dagger={\cal P}c_\alpha^\dagger{\cal P}\,.
\label{graded}
\end{align} 
We note in passing that the graded dagger operation on the matrices given in (\ref{graded}) reads
\begin{align}
 {\tilde b}_\alpha^\ddagger={\tilde b}_\alpha^\dagger\,,\quad ({\tilde b}_\alpha^\dagger)^\ddagger = - {\tilde b}_\alpha \,\quad {\tilde c}_\alpha^\ddagger={\tilde c}_\alpha^\dagger\,,\quad ({\tilde c}_\alpha^\dagger)^\ddagger = - {\tilde c}_\alpha \,.
\end{align}
Finally, in (\ref{LambdaL}) and (\ref{LambdaR}), it is understood that the columns and rows of zero are deleted after the projection and therefore, we have $9\times9$ matrices $(\Lambda^L_M, \Lambda^R_M)$ as intended.

\section{Conclusions}

In this paper we have a studied a particular deformation of the $N=4$ SYM theory with cubic SSB and mass deformation terms. We have determined a family of fuzzy vacua which are expressed in terms of direct sums of product of two fuzzy spheres. Structure of these vacuum configurations is revealed by permitting splittings of the scalar fields that involve the introduction of $k_1+k_2$ component multiplets transforming under the representation $(\frac{k_1-1}{2},0)\oplus (0,\frac{k_2-1}{2})$ of the global symmetry and it is found that all fuzzy monopole sectors over $S_F^2 \times S_F^2$ are systematically accessed thorough projections of these vacua. Focusing on the simplest member $S_F^{2\, Int}\times S_F^{2\, Int}$ of this family, we have demonstrated that the fluctuations about this vacuum have precisely the form of gauge fields, which allowed us to conjecture that the emerging model is an effective $U(n)$ $(n < {\cal N})$ gauge theory on $M^4 \times S_F^{2\, Int}\times S_F^{2\, Int}$. To support this interpretation, we have studied the $U(4)$ model and obtained all the $SU(2)\times SU(2)$-equivariant fields, which characterized its low energy degrees of freedom and also examined the monopole sectors with winding numbers $(\pm 1,0),\,(0,\pm1),\,(\pm1,\pm 1)$ in some detail. We have noted that spinorial modes that naturally come out of this analysis do not comprise independent degrees of freedom in the effective theory, but they may be used to find the "square roots" of the equivariant gauge field modes. Finally, we have seen that  $S_F^{2\, Int}\times S_F^{2\, Int}$ identifies with the bosonic part of the product of two fuzzy superspheres with $OSP(2,2)\times OSP(2,2)$ supersymmetry and discussed how it comes. We would like to stress that our results applies just as well to Yang Mills matrix models with the same type of vacua and and methods are quite versatile to investigate other fuzzy vacuum configurations, which may be of physical interest.

\vskip 1em

\appendices

\section{Some Details for Sections $2$ and $3$}

Variation of the action \eqref{action} with respect to $\Phi_a^{i \, L}$ gives 
\be
D_\mu D^\mu \Phi_a^{i \, L}  + \frac{1}{g_L^2} ( 2 f_{ijk} \Phi_b^{j \, L} F_{ab}^{k \, L} - \varepsilon_{abc} F_{bc}^{i \, L} ) = 0 \,, 
\ee
while the variation with respect to $\Psi_\alpha^{l \, L \, \dagger}$ yields
\be
\left ( D_\mu D^\mu \Phi_a^{i \, L}  + \frac{1}{g_L^2} ( 2 f_{ijk} \Phi_b^{j \, L} F_{ab}^{k \, L} - \varepsilon_{abc} F_{bc}^{i \, L} ) \right ) \gamma_{lmi}( \tilde{\tau}_a \Psi^{m \, L})_\alpha = 0 \,, 
\ee
where $\Phi_a^{L} = \Phi_a^{i \, L} \lambda_{i}$, $\Psi_\alpha^{L} = \Psi_\alpha^{i \, L} \lambda_{i}$ with the anti-hermitian $SU({\cal N})$ generators $\lambda_i$ $(i = 1 \, \cdots \,, {\cal N}^2-1)$ fulfilling $\lambda_i \lambda_j = -\frac{2}{{\cal N}} \delta_{ij} + (d_{ijk} + f_{ijk})\lambda_k$ and $\gamma_{ijk} := d_{ijk} + f_{ijk}$ for short. Clearly, these equations imply each other. Variation with respect to  $\Phi_a^{i \, R}$ and $\Psi_\alpha^{l \, R \, \dagger}$ yield analogous expressions with $L \rightarrow R$.

The block diagonal form $({\cal D}_a^L,{\cal D}_a^R)$ indicated in page $9$ is given as 
\begin{multline}
 {\cal D}_a^L{\cal D}_a^L+{\cal D}_a^R{\cal D}_a^R=\bigg(-\big(\ell_L(\ell_L+1)+\ell_R(\ell_R+1)\big)\bm{1}_{(2\ell_L+1)(2\ell_R+1)4n},
 \\-\big((\ell_L-\frac{1}{2})(\ell_L+\frac{1}{2})+\ell_R(\ell_R+1)\big)\bm{1}_{(2\ell_L)(2\ell_R+1)2n},-\big((\ell_L+\frac{1}{2})(\ell_L+\frac{3}{2})+\ell_R(\ell_R+1)\big)\bm{1}_{(2\ell_L+2)(2\ell_R+1)2n},
 \\-\big(\ell_L(\ell_L+1)+(\ell_R-\frac{1}{2})(\ell_R+\frac{1}{2})\big)\bm{1}_{(2\ell_L+1)(2\ell_R)2n},-\big(\ell_L(\ell_L+1)+(\ell_R+\frac{1}{2})(\ell_R+\frac{3}{2})\big)\bm{1}_{(2\ell_L+1)(2\ell_R+2)2n},
 \\-\big((\ell_L-\frac{1}{2})(\ell_L+\frac{1}{2})+(\ell_R-\frac{1}{2})(\ell_R+\frac{1}{2})\big)\bm{1}_{(2\ell_L)(2\ell_R)n},
 \\-\big((\ell_L+\frac{1}{2})(\ell_L+\frac{3}{2})+(\ell_R-\frac{1}{2})(\ell_R+\frac{1}{2})\big)\bm{1}_{(2\ell_L+2)(2\ell_R)n},
 \\-\big((\ell_L-\frac{1}{2})(\ell_L+\frac{1}{2})+(\ell_R+\frac{1}{2})(\ell_R+\frac{3}{2})\big)\bm{1}_{(2\ell_L)(2\ell_R+2)n}, 
 \\-\big((\ell_L+\frac{1}{2})(\ell_L+\frac{3}{2})+(\ell_R+\frac{1}{2})(\ell_R+\frac{3}{2})\big)\bm{1}_{(2\ell_L+2)(2\ell_R+2)n}\bigg) \,.
 \label{calD}
\end{multline}

The matrices in \eqref{i1} and  \eqref{i2} square as 
\begin{equation}
 \begin{split}
  &(Q^L_B)^2=-\bm {1}_{(2\ell_L+1)(2\ell_R+1)64}\,,\quad  (Q^R_B)^2=-\bm {1}_{(2\ell_L+1)(2\ell_R+1)64}\,,\quad (Q^L_{\pm})^2=-\Pi_{\pm}^L\,,
  \\&(Q^R_{\pm})^2=-\Pi_{\pm}^R\,,\quad {(Q_{0_{\bm 0}}^L)}^2=-{\Pi}_{0_{\bm 0}}^L\,,\quad {(Q_{0_{\bm 0}}^R)}^2=-{\Pi}_{0_{\bm 0}}^R\,,\quad {(Q_{0_{\bm 2}}^L)}^2=-{\Pi}_{0_{\bm 2}}^L\,,\quad {(Q_{0_{\bm 2}}^R)}^2=-{\Pi}_{0_{\bm 2}}^R\,,
  \\&{(iS_i^L)}^2=-\Pi_0^L\,,\quad {(iS_i^R)}^2=-\Pi_0^R\,,\quad ({Q_{S_i}^L})^2=-\Pi_0^L\,,\quad ({Q_{S_i}^R})^2=-\Pi_0^R\,,\quad (Q_F^L)^2=-\Pi_{\frac{1}{2}}^L\,,
  \\&(Q_F^R)^2=-\Pi_{\frac{1}{2}}^R\,,\quad  (Q_H^L)^2=-\Pi_{\frac{1}{2}}^L\,, \quad  (Q_H^R)^2=-\Pi_{\frac{1}{2}}^R\,,\quad (Q^L_{BI})^2=-\Pi_{\frac{1}{2}}^L\,,\quad (Q^R_{BI})^2=-\Pi_{\frac{1}{2}}^R\,, \\&(Q^L_{I})^2=-\Pi_{\frac{1}{2}}^L\,, \quad (Q^R_{I})^2=-\Pi_{\frac{1}{2}}^R\,,
 \end{split}
 \end{equation}
justifying that they are ``idempotent''s in the subspace the belong to.

Using the equivariant invariants in (\ref{196}), vectors in the $(1,0)$ IRR may be listed as
\begin{align}
 &\Pi_i^R[D_a^L,Q_j^L],\quad\Pi_i^RQ_j^L[D_a^L,Q_j^L],\quad \Pi_i^R\lbrace D_a^L, Q_j^L\rbrace,\quad S_k^R[D_a^L,Q_j^L],\quad S_k^RQ_j^L[D_a^L,Q_j^L],\nonumber
 \\&S_k^R\lbrace D_a^L, Q_j^L\rbrace\quad Q_j^R[D_a^L,Q_j^L],\quad Q_j^RQ_j^L[D_a^L,Q_j^L],\quad Q_j^R\lbrace D_a^L, Q_j^L\rbrace\quad Q_{S_k}^R[D_a^L,Q_j^L],\nonumber
 \\& Q_{S_k}^RQ_j^L[D_a^L,Q_j^L],\quad Q_{S_k}^R\lbrace D_a^L, Q_j^L\rbrace,\quad \Pi_i^R[D_a^L,Q_{S_k}^L],\quad\Pi_i^RQ_0^L[D_a^L,Q_{S_k}^L],\quad \Pi_i^R\lbrace D_a^L, Q_{S_k}^L\rbrace,\nonumber
 \\& S_k^R[D_a^L,Q_{S_k}^L],\quad S_k^RQ_0^L[D_a^L,Q_{S_k}^L],\quad S_k^R\lbrace D_a^L, Q_{S_k}^L\rbrace\quad Q_j^R[D_a^L,Q_{S_k}^L],\quad Q_j^RQ_0^L[D_a^L,Q_{S_k}^L],\nonumber
 \\& Q_j^R\lbrace D_a^L, Q_{S_k}^L\rbrace,\quad Q_{S_k}^R[D_a^L,Q_{S_k}^L],\quad Q_{S_k}^RQ_0^L[D_a^L,Q_{S_k}^L],\quad Q_{S_k}^R\lbrace D_a^L,Q_{S_k}^L\rbrace,\quad \Pi_i^R\Pi_i^L\omega_a^L,\nonumber
 \\&S_k^R\Pi_i^L\omega_a^L\quad Q_j^R\Pi_i^L\omega_a^L\quad Q_{S_k}^R\Pi_i^L\omega_a^L,\quad \Pi_i^RS_k^L\omega_a^L,\quad S_k^RS_k^L\omega_a^L\quad Q_j^RS_k^L\omega_a^L\quad Q_{S_k}^RS_k^L\omega_a^L,
 \label{vect}
\end{align} where $Q_0^L=Q_{0_{\bm 0}}^L+Q_{0_{\bm 2}}^L$. Equivariant vectors in the $(0,1)$ IRR is obtained from (\ref{vect}) simply by the exchange $ L\leftrightarrow R$. 

$336$ equivariant spinors in the IRR $(\frac{1}{2},0)$ parametrized as
\begin{align}
 &\Pi_i^R\Pi_\mu^L\beta_{\alpha}^LQ_{\nu}^L,\quad \Pi_i^R\Pi_\nu^L\beta_{\alpha}^LQ_{\mu}^L,\quad \Pi_i^RQ_\mu^L\beta_{\alpha}^L\Pi_{\nu}^L,\quad \Pi_i^RQ_\nu^L\beta_{\alpha}^L\Pi_{\mu}^L,\quad \Pi_i^RQ_\mu^L\beta_{\alpha}^LQ_{\nu}^L,\quad\Pi_i^RQ_\nu^L\beta_{\alpha}^LQ_{\mu}^L\nonumber
 \\&\Pi_i^RS_\rho^L\beta_{\alpha}^L\Pi_{\nu}^L,\quad\Pi_i^R\Pi_\nu^L\beta_{\alpha}^LS_{\rho}^L,\quad\Pi_i^RQ_{S_\rho}^L\beta_{\alpha}^L\Pi_{\nu}^L,\quad\Pi_i^R\Pi_\nu^L\beta_{\alpha}^LQ_{S_\rho}^L,\quad \Pi_i^RQ_{S_\rho}^L\beta_{\alpha}^LQ_{\nu}^L,\quad \Pi_i^RQ_\nu^L\beta_{\alpha}^LQ_{S_\rho}^L\nonumber
 \\&S_k^R\Pi_\mu^L\beta_{\alpha}^LQ_{\nu}^L,\quad S_k^R\Pi_\nu^L\beta_{\alpha}^LQ_{\mu}^L,\quad S_k^RQ_\mu^L\beta_{\alpha}^L\Pi_{\nu}^L,\quad S_k^RQ_\nu^L\beta_{\alpha}^L\Pi_{\mu}^L,\quad S_k^RQ_\mu^L\beta_{\alpha}^LQ_{\nu}^L,\quad S_k^RQ_\nu^L\beta_{\alpha}^LQ_{\mu}^L\nonumber
 \\&S_k^RS_\rho^L\beta_{\alpha}^L\Pi_{\nu}^L,\quad S_k^R\Pi_\nu^L\beta_{\alpha}^LS_{\rho}^L,\quad S_k^RQ_{S_\rho}^L\beta_{\alpha}^L\Pi_{\nu}^L,\quad S_k^R\Pi_\nu^L\beta_{\alpha}^LQ_{S_\rho}^L,\quad S_k^RQ_{S_\rho}^L\beta_{\alpha}^LQ_{\nu}^L,\quad S_k^RQ_\nu^L\beta_{\alpha}^LQ_{S_\rho}^L,\nonumber
 \\&Q_j^R\Pi_\mu^L\beta_{\alpha}^LQ_{\nu}^L,\quad Q_j^R\Pi_\nu^L\beta_{\alpha}^LQ_{\mu}^L,\quad Q_j^RQ_\mu^L\beta_{\alpha}^L\Pi_{\nu}^L,\quad Q_j^RQ_\nu^L\beta_{\alpha}^L\Pi_{\mu}^L,\quad Q_j^RQ_\mu^L\beta_{\alpha}^LQ_{\nu}^L,\quad Q_j^RQ_\nu^L\beta_{\alpha}^LQ_{\mu}^L\nonumber
 \\&Q_j^RS_\rho^L\beta_{\alpha}^L\Pi_{\nu}^L,\quad Q_j^R\Pi_\nu^L\beta_{\alpha}^LS_{\rho}^L,\quad Q_j^RQ_{S_\rho}^L\beta_{\alpha}^L\Pi_{\nu}^L,\quad Q_j^R\Pi_\nu^L\beta_{\alpha}^LQ_{S_\rho}^L,\quad Q_j^RQ_{S_\rho}^L\beta_{\alpha}^LQ_{\nu}^L,\quad Q_j^RQ_\nu^L\beta_{\alpha}^LQ_{S_\rho}^L\nonumber
 \\&Q_{S_k}^R\Pi_\mu^L\beta_{\alpha}^LQ_{\nu}^L,\quad Q_{S_k}^R\Pi_\nu^L\beta_{\alpha}^LQ_{\mu}^L,\quad Q_{S_k}^RQ_\mu^L\beta_{\alpha}^L\Pi_{\nu}^L,\quad Q_{S_k}^RQ_\nu^L\beta_{\alpha}^L\Pi_{\mu}^L,\quad Q_{S_k}^RQ_\mu^L\beta_{\alpha}^LQ_{\nu}^L,\quad Q_{S_k}^RQ_\nu^L\beta_{\alpha}^LQ_{\mu}^L\nonumber
 \\&Q_{S_k}^RS_\rho^L\beta_{\alpha}^L\Pi_{\nu}^L,\quad Q_{S_k}^R\Pi_\nu^L\beta_{\alpha}^LS_{\rho}^L,\quad Q_{S_k}^RQ_{S_\rho}^L\beta_{\alpha}^L\Pi_{\nu}^L,\quad Q_{S_k}^R\Pi_\nu^L\beta_{\alpha}^LQ_{S_\rho}^L,\quad Q_{S_k}^RQ_{S_\rho}^L\beta_{\alpha}^LQ_{\nu}^L,\quad Q_j^RQ_\nu^L\beta_{\alpha}^LQ_{S_\rho}^L \,,
 \label{sspi}
\end{align} 
where $\beta_{\alpha}^L= {\bm 1}^{2 \ell_L+1} \otimes  {\bm 1}^{2 \ell_R+1}\otimes b_{\alpha} \otimes {\bm 1}_4,\,\,\,\beta_{\alpha}^R= {\bm 1}^{2 \ell_L+1} \otimes  {\bm 1}^{2 \ell_R+1}\otimes c_{\alpha} \otimes {\bm 1}_4,\quad \mu=0_{\bm 0},0_{\bm 2},\quad \nu=+,-,\quad \rho=1,2$ and where $\Pi_{0_{\bm 0}}^L,Q_{0_{\bm 0}}^L, S_1^L, Q_{S_1}^L$ on the left most and $\Pi_{0_{\bm 2}}^L,Q_{0_{\bm 2}}^L, S_2^L, Q_{S_2}^L$ on the right most side in any of these expressions are excluded. For the equivariant spinors carrying $(0,\frac{1}{2})$ representation, it is enough to take $L\leftrightarrow R$ in (\ref{sspi}).

\section{Some Details for Sections $4$}

The vacuum configuration with $(k_1\,,k_2)$ component multiplets can be calculated for the cases $k_1=\text{even}\,,k_2=\text{even}$ and $k_1=\text{even}\,,k_2=\text{odd}$ as follows

\begin{align}
  {S_F^{2\,Int}}&_{k_1\,even} \times {S_F^{2\,Int}}_{k_2\,even}:=\nonumber \\ 
  & S_F^2(\ell_L+\ell_{\frac{k_1}{2}}^{k_1})\times S_F^2(\ell_R+\ell_{\frac{k_2}{2}}^{k_2})\oplus\cdots \oplus S_F^2(\ell_L+\ell_{\frac{k_1}{2}}^{k_1})\times S_F^2(|\ell_R-\ell_{\frac{k_2}{2}}^{k_2}|)\nonumber 
  \\\oplus& \vdots\quad \nonumber
  \\ \oplus &S_F^2(|\ell_L-\ell_{\frac{k_1}{2}}^{k_1}|)\times S_F^2(\ell_R+\ell_{\frac{k_2}{2}}^{k_2})\oplus \cdots \oplus S_F^2(|\ell_L-\ell_{\frac{k_1}{2}}^{k_1}|)\times S_F^2(|\ell_R-\ell_{\frac{k_2}{2}}^{k_2}|)\nonumber
  \\\oplus &\bm{4}\sum_{n=0}^{\frac{k_1}{2}-1}\sum_{m=0}^{\frac{k_2}{2}-1}\bigg[S_F^2(\ell_L+\ell^{k_1}_n)\times S_F^2(\ell_R+\ell^{k_2}_m)\oplus \cdots \oplus S_F^2(\ell_L+\ell^{k_1}_n)\times S_F^2(|\ell_R-\ell^{k_2}_m|)\nonumber
 \\\oplus &\vdots \nonumber
 \\\oplus & S_F^2(|\ell_L-\ell^{k_1}_n|)\times S_F^2(\ell_R+\ell^{k_2}_m)\oplus \cdots \oplus S_F^2(|\ell_L-\ell^{k_1}_n|)\times S_F^2(|\ell_R-\ell^{k_2}_m|)\bigg]\nonumber
 \\\oplus &\bm{2}\sum_{n=0}^{\frac{k_1}{2}-1}\bigg[S_F^2(\ell_L+\ell^{k_1}_n)\times S_F^2(\ell_R+\ell^{k_2}_{\frac{k_2}{2}})\oplus \cdots \oplus S_F^2(\ell_L+\ell^{k_1}_n)\times S_F^2(|\ell_R-\ell^{k_2}_{\frac{k_2}{2}}|)\nonumber
 \\\oplus & \vdots\nonumber
 \\\oplus & S_F^2(|\ell_L-\ell^{k_1}_n|)\times S_F^2(\ell_R+\ell^{k_2}_{\frac{k_2}{2}})\oplus \cdots \oplus S_F^2(|\ell_L-\ell^{k_1}_n|)\times S_F^2(|\ell_R-\ell^{k_2}_{\frac{k_2}{2}}|)\bigg]\nonumber
 \\\oplus & \bm{2}\sum_{m=0}^{\frac{k_2}{2}-1}\bigg[S_F^2(\ell_L+\ell^{k_1}_{\frac{k_1}{2}})\times S_F^2(\ell_R+\ell^{k_2}_m)\oplus \cdots \oplus S_F^2(\ell_L+\ell^{k_1}_{\frac{k_1}{2}})\times S_F^2(|\ell_R-\ell^{k_2}_m|)\nonumber
 \\\oplus & \vdots\nonumber
 \\\oplus & S_F^2(|\ell_L-\ell^{k_1}_{\frac{k_1}{2}}|)\times S_F^2(\ell_R+\ell^{k_2}_m)\oplus \cdots \oplus S_F^2(|\ell_L-\ell^{k_1}_{\frac{k_1}{2}}|)\times S_F^2(|\ell_R-\ell^{k_2}_m|)\bigg]\,.
 \label{geneven}
\end{align}

\begin{align}
 {S_F^{2\,Int}}&_{k_1\,even}  \times {S_F^{2\,Int}}_{k_2\,odd}:=\nonumber \\
 \bm{4} & \sum_{n=0}^{\frac{k_1}{2}-1}\sum_{m=0}^{\frac{k_2-1}{2}}\bigg [S_F^2(\ell_L+\ell^{k_1}_n)\times S_F^2(\ell_R+\ell^{k_2}_m)\oplus \cdots \oplus S_F^2(\ell_L+\ell^{k_1}_n)\times S_F^2(|\ell_R-\ell^{k_2}_m|)\nonumber
 \\\oplus & S_F^2(\ell_L+\ell^{k_1}_n-1)\times S_F^2(\ell_R+\ell^{k_2}_m)\oplus \cdots \oplus S_F^2(\ell_L+\ell^{k_1}_n-1)\times S_F^2(|\ell_R-\ell^{k_2}_m|)\nonumber
 \\\oplus & \vdots \nonumber
 \\\oplus &S_F^2(|\ell_L-\ell^{k_1}_n|)\times S_F^2(\ell_R+\ell^{k_2}_m)\oplus \cdots \oplus S_F^2(|\ell_L-\ell^{k_1}_n|)\times S_F^2(|\ell_R-\ell^{k_2}_m|)\bigg]\nonumber
 \\\oplus & \bm{2}\sum_{m=0}^{\frac{k_2-1}{2}}\bigg[S_F^2(\ell_L+\ell^{k_1}_{\frac{k_1}{2}})\times S_F^2(\ell_R+\ell^{k_2}_m)\oplus \cdots \oplus S_F^2(\ell_L+\ell^{k_1}_{\frac{k_1}{2}})\times S_F^2(|\ell_R-\ell^{k_2}_m|)\nonumber
 \\\oplus & \vdots\nonumber
 \\\oplus & S_F^2(|\ell_L-\ell^{k_1}_{\frac{k_1}{2}}|)\times S_F^2(\ell_R+\ell^{k_2}_m)\oplus \cdots \oplus S_F^2(|\ell_L-\ell^{k_1}_{\frac{k_1}{2}}|)\times S_F^2(|\ell_R-\ell^{k_2}_m|)\bigg]\,.
 \label{genevenodd}
\end{align}

\section{Another Vacuum Solution}

It is worthwhile to ask whether it is possible to find solutions to equations given in (\ref{minpot}) in the form
\begin{equation}
\begin{split}
  \Phi_a^L&=(X_a^{(2\ell_L+1)}\otimes \bm{1}^{(2\ell_R+1)}\otimes \bm{1}_4\otimes \bm{1}_n)+(\bm{1}^{(2\ell_L+1)}\otimes \bm{1}^{(2\ell_R+1)}\otimes \tilde{\Gamma_a^0}^L\otimes \bm{1}_n )\,,
  \\\Phi_a^R&=(\bm{1}^{(2\ell_L+1)}\otimes X_a^{(2\ell_R+1)}\otimes \bm{1}_4\otimes \bm{1}_n)+(\bm{1}^{(2\ell_L+1)}\otimes \bm{1}^{(2\ell_R+1)}\otimes \tilde{\Gamma_a^0}^R\otimes \bm{1}_n )\,,
  \label{vacuum}
 \end{split}
 \end{equation}
with the factorization ${\cal N}=(2\ell_L+1)\times(2\ell_R+1)\times 4\times n$ and where  $\tilde{\Gamma_a^0}^L$ and $\tilde{\Gamma_a^0}^R$ are $4\times 4$ matrices instead of the $16\times 16$ matrices determined in section \ref{section2.2}, satisfying the relations in (\ref{SU(2)com}). The answer to this question is only superfically affirmative as such $\tilde{\Gamma_a^0}^L$ and $\tilde{\Gamma_a^0}^R$ exist, but against the very premise of our initial requirement that $\tilde{\Gamma_a^0}^L$ and $\tilde{\Gamma_a^0}^R$ are bilinears of the doublets $\Psi^L$ and $\Psi^R$ of $SU(2)\times SU(2)$ transforming under its $(\frac{1}{2},0)$ and $(0,\frac{1}{2})$ IRR's. To be more concrete, it turns out that it is possible to express $\tilde{\Gamma_a^0}^L$ and $\tilde{\Gamma_a^0}^R$ in terms of bilinears of some matrices $\chi^L$ and $\chi^R$, which, however, do not transform as $(\frac{1}{2},0)$ and $(0,\frac{1}{2})$ under $SU(2)\times SU(2)$. This fact suggests that, we should expect to find no equivariant spinor field modes at all for the emerging effective $U(4)$ gauge theory. It appears instructive to examine this case in some detail.
 
If we start with two sets of fermionic annihilation-creation operators $d_\alpha, d_\alpha^\dagger$ with
 \begin{align}
 &\lbrace d_{\alpha},d_{\beta}\rbrace=0\,,\quad \lbrace d_{\alpha}^\dagger,d_{\beta}^\dagger\rbrace=0\,,\quad \lbrace d_{\alpha},d_{\beta}^\dagger\rbrace=\delta_{\alpha \beta}\,,\quad \alpha,\beta=1,2
 \end{align}  which span the four-dimensional Hilbert space,
 \begin{align}
  |n_1,\,n_2\rangle \equiv (d_1^\dagger)^{n_1}(d_2^\dagger)^{n_2}|0,\,0\rangle,\quad n_1,n_2=0,1\,,
  \label{hilbert}
 \end{align}
and choose the two-component objects 
\begin{equation}
 \chi^L = \left (
\begin{array}{c}
\chi^L_1 \\
\chi^L_2
\end{array}
\right )
:=
\left (
\begin{array}{c}
d_1 \\
d_2
\end{array}
\right ) \,,\quad
 {\chi}^R = \left (
\begin{array}{c}
{\chi^R_1} \\
{\chi^R_2}
\end{array}
\right )
:=
\left (
\begin{array}{c}
d_1^\dagger \\
d_2
\end{array}
\right ) \,,
\end{equation} then, $\tilde{\Gamma_a^0}^L=-\frac{i}{2}{\chi^L}^\dagger\tau_a\chi^L\,, {\Gamma_a^0}^R=-\frac{i}{2}{\chi^R}^\dagger\tau_a\chi^R$ satisfy
\begin{align}
&[\tilde{\Gamma_a^0}^L,\tilde{\Gamma_b^0}^L]=\epsilon_{abc}\tilde{\Gamma_c^0}^L\,,\quad [\tilde{\Gamma_a^0}^R,\tilde{\Gamma_b^0}^R]=\epsilon_{abc}\tilde{\Gamma_c^0}^R\,,\quad [\tilde{\Gamma_a^0}^L,\tilde{\Gamma_b^0}^R]=0\,.
\label{gamma}
\end{align} However, we find that
\begin{align}
 &[\chi^L_{\alpha},\tilde{\Gamma_a^0}^L]=-\frac{i}{2}(\tau_a)_{\alpha \beta}\chi^L_{\beta}\,,\quad [\chi^L_{\alpha},\tilde{\Gamma_a^0}^R]\neq 0\,,\quad [\chi^R_{\alpha},\tilde{\Gamma_a^0}^R]=-\frac{i}{2}(\tau_a)_{\alpha \beta}\chi^R_{\beta}\,,\quad [\chi^R_{\alpha},\tilde{\Gamma_a^0}^L]\neq 0\,.
 \label{nonvanis}
 \end{align} Thus, due to the two nonvanishing commutators in (\ref{nonvanis}), $\chi^L$ and $\chi^R$ are \textit{not} transforming 
in the IRRs $(\frac{1}{2},0)$ and $(0, \frac{1}{2})$ of $SU(2)\times SU(2)$, respectively. Bearing this fact in mind, we can nevertheless continue to work with the matrices $\tilde{\Gamma_a^0}^L$ and $\tilde{\Gamma_a^0}^R$ satisfying (\ref{gamma}), and investigate the structure of the emerging model in its own right.

 Using the identities
  \begin{equation}
  {(\tilde{\Gamma_a^0}^L})^2=-\frac{3}{4}N+\frac{3}{2}N_1N_2\,,\quad {(\tilde{\Gamma_a^0}^R})^2=\frac{3}{4}N-\frac{3}{2}N_1N_2-\frac{3}{4}\,,
  \label{casimir}
 \end{equation}where $N=N_1+N_2,\,\,N_1=b^\dagger_1b_1,\,\, N_2=b^\dagger_2b_2$, the quadratic Casimir operator can be evaluated and we simply find 
 \begin{align}
  C_2={(\tilde{\Gamma_a^0}^L})^2+{(\tilde{\Gamma_a^0}^R})^2=-\frac{3}{4}\mathbf{1}_4\,.
 \end{align} This means that $(\tilde{\Gamma_a^0}^L,\tilde{\Gamma_a^0}^R)$ carry the direct sum representation  $(\frac{1}{2},0)\oplus (0,\frac{1}{2})$.

The Hilbert space in (\ref{hilbert}) has four states: $|0,\,0\rangle,\,\,|0,\,1\rangle,\,\,|1,\,0\rangle,\,\,|1,\,1\rangle$. $\tilde{\Gamma_a^0}^L$ is reducible with respect to $SU(2)_L$ and has two inequivalent singlets, $|0,\,0\rangle,\,\,|1,\,1\rangle$ and a doublet, spanned by $|0,\,1\rangle,\,\,|1,\,0\rangle$. Similarly,  $\tilde{\Gamma_a^0}^R$ is reducible with respect to $SU(2)_R$ and has two inequivalent singlets, $|0,\,1\rangle,\,\,|1,\,0\rangle$, and a doublet, spanned by $|0,\,0\rangle,\,\,|1,\,1\rangle$:
\begin{equation}
\begin{split}
&\tilde{\Gamma_a^0}^L \rightarrow (0_{\bm {0}},0)\oplus (0_{\bm{2}},0) \oplus (\frac{1}{2},0)\,,
\\&\tilde{\Gamma_a^0}^R \rightarrow (0,0_{\bm {0}})\oplus (0,0_{\bm{2}}) \oplus (0,\frac{1}{2})\,.
\end{split}
\end{equation}
Two inequivalent singlets of $\tilde{\Gamma_a^0}^L$ can be distinguished by the eigenvalues $0,2$ of $N$, since $[\tilde{\Gamma_a^0}^L,N]=0$. Likewise, the eigenvalues $0,2$ of the operator $(\bm{1}_4-(N_1-N_2))$ distinguishes the two inequivalent singlets of $\tilde{\Gamma_a^0}^R$ since $[\tilde{\Gamma_a^0}^R,\bm{1}_4-(N_1-N_2)]=0$.

Let us define the two projectors 
\begin{equation}
\begin{split}
 &P_0=\frac{{(\tilde{\Gamma_a^0}^L})^2+\frac{3}{4}}{\frac{3}{4}}=-\frac{{(\tilde{\Gamma_a^0}^R})^2}{\frac{3}{4}}=1-N+2N_1N_2\,,
 \\&P_{\frac{1}{2}}=-\frac{{(\tilde{\Gamma_a^0}^L})^2}{\frac{3}{4}}=\frac{{(\tilde{\Gamma_a^0}^R})^2+\frac{3}{4}}{\frac{3}{4}}=N-2N_1N_2\,,
\end{split}
\end{equation}
where $P_0$ projects to the singlets of ${\tilde{\Gamma_a^0}^L}$ and to the doublet of ${\tilde{\Gamma_a^0}^R}$, and $P_{\frac{1}{2}}$ projects to the doublet of ${\tilde{\Gamma_a^0}^L}$ and to the singlet of ${\tilde{\Gamma_a^0}^R}$. 
Projections to the inequivalent singlets and spin up and down components of doublets read
\begin{equation}
\label{proLR}
\begin{split}
&P^L_{0_{\bm 0}}=-\frac{1}{2}(N-2)P_0=1-N+N_1N_2\,,\quad P^L_{0_{\bm 2}}=\frac{1}{2}NP_0=N_1N_2\,,
\\&P^L_{\frac{1}{2}+}=P_{\frac{1}{2}}N_1=N_1-N_1N_2\,,\quad P^L_{\frac{1}{2}-}=P_{\frac{1}{2}}N_2=N_2-N_1N_2\,,
\\&{P}_{0_{\bm 0}}^R=P^L_{\frac{1}{2}+}\,,\quad {P}_{0_{\bm 2}}^R=P^L_{\frac{1}{2}-}\,,\quad P^R_{\frac{1}{2}+}=P^L_{0_{\bm 0}}\,\quad P^R_{\frac{1}{2}-}=P^L_{0_{\bm 2}}\,.
\end{split}
\end{equation}

The Clebsch-Gordan decomposition of the vacuum configuration proposed in equation (\ref{vacuum}) is determined as
\begin{align}
  &(\ell_L,\ell_R)\otimes\left((\frac{1}{2},0)\oplus(\frac{1}{2},0)\right) \equiv (\ell_L+\frac{1}{2},\ell_R)\oplus (\ell_L-\frac{1}{2},\ell_R)\oplus (\ell_L,\ell_R+\frac{1}{2})\oplus (\ell_L,\ell_R-\frac{1}{2})\,. 
  \label{concentricfuzzy}
\end{align}
This means that the vacuum configuration can be written as the direct sum
\begin{multline}
 {S_F^{2\,Int}}\times {S_F^{2\,Int}}\equiv \left (S_F^2(\ell_L+\frac{1}{2})\times S_F^2(\ell_R)\right)\oplus \left (S_F^2(\ell_L-\frac{1}{2})\times S_F^2(\ell_R)\right)
 \\\oplus\left(S_F^2(\ell_L)\times S_F^2(\ell_R+\frac{1}{2})\right)
 \oplus \left(S_F^2(\ell_L)\times S_F^2(\ell_R-\frac{1}{2})\right)\,.
 \label{configure}
\end{multline}
Projections to each summand in (\ref{configure}) can be obtained by adapting the formula in(\ref{projector}) to the present case. This yields the projectors $\Pi_{\alpha\beta}\equiv \lbrace\Pi_{+ 0}\,,\Pi_{- 0}\,,\Pi_{0 +}\,,\Pi_{0 -}\rbrace$ (see, equation (\ref{ppp}) below)
which, upon using the suitably adapted version of (\ref{projector2}), are unitarily equivalent to the product $\Pi_\alpha^L\Pi_\beta^R$, which we write as $\Pi_{\alpha\beta}\equiv\Pi_\alpha^L\Pi_\beta^R$.

For the projectors $\Pi^L_0\,,\Pi^R_0\,,\Pi^L_{\pm}\,,\Pi^R_{\pm}$, we have the explicit forms 
 \begin{align}
 &\Pi^L_0= \bm{1}^{(2\ell_L+1)}\otimes \bm{1}^{(2\ell_R+1)}\otimes P_{0}\otimes \bm{1}_n\,,\quad \Pi^R_0= \bm{1}^{(2\ell_L+1)}\otimes \bm{1}^{(2\ell_R+1)}\otimes P_{\frac{1}{2}}\otimes \bm{1}_n\,,\nonumber
  \\&\Pi_{\pm}^L=\frac{1}{2}(\pm i{ Q}^L_I +\Pi^L_{\frac{1}{2}})\,,\quad\Pi_{\pm}^R = \frac{1}{2} (\pm i {Q}^R_I + \Pi^R_{\frac{1}{2}})\,,
\end{align}
where 
\begin{align}
{Q}^L_I=i\frac{X_a^L\tilde{\Gamma_a^0}^L-\frac{1}{4}\Pi^L_{\frac{1}{2}}}{\frac{1}{2}(\ell_L+\frac{1}{2})}\,,\quad { Q}^R_I=i\frac{X_a^R\tilde{\Gamma_a^0}^R-\frac{1}{4}\Pi^R_{\frac{1}{2}}}{\frac{1}{2}(\ell_R+\frac{1}{2})}\,.
\end{align} In observation of the relations given in (\ref{proLR}), we see that 
\begin{equation}
 \Pi_{\pm 0}\equiv \Pi_{\pm}^L\Pi^R_0=\Pi_{\pm}^L\,,\quad \Pi_{0 \pm}\equiv \Pi_{0}^L\Pi^R_{\pm}=\Pi_{\pm}^L\,,
 \label{ppp}
\end{equation}
while all other products vanish. Therefore, $\Pi_{\pm}^R\,,\Pi_{\pm}^L$ are simply the required four projectors. For convenience, we list them in the table below. 
 \begin{center}
    \begin{tabular}{c | c }
    Projector & To the Representation \\ \hline
    \\
    $\Pi_{\pm}^L = \frac{1}{2} (\pm i { Q}^L_I + \Pi^L_{\frac{1}{2}})$ & $(\ell_L \pm \frac{1}{2},\ell_R)$ \\
    $\Pi_{\pm}^R = \frac{1}{2} (\pm i { Q}^R_I + \Pi^R_{\frac{1}{2}})$ & $(\ell_L,\ell_R \pm \frac{1}{2})$
\end{tabular}
\end{center}
 
 At this stage we can consider the fluctuations about the vacuum configuration (\ref{vacuum})
 \begin{equation}
  \label{fluctuation}
 \begin{split}
  &\Phi_a^L=X_a^L+\tilde{\Gamma_a^0}+A^L_a:=D_a^L+A_a^L\,,
  \\& \Phi_a^R=X_a^R+\tilde{\Gamma_a^0}+A^R_a:=D_a^R+A_a^R\,,
 \end{split}
 \end{equation}
 where $A_a^L,\,A_a^R\in u(2\ell_L+1)\otimes u(2\ell_R+1)\otimes u(4)\otimes u(n)$. 
 
 We can view $A_a^L$ and $A_a^R$ ($a=1,2,3$) as the six components of a $U(n)$ gauge field on $S_F^{2\,Int}\times S_F^{2\,Int}$ since $F_{ab}^L,\,F_{ab}^R,\,F_{ab}^{L,R}$ take the form of the curvature tensor
 \begin{equation}
\begin{split}
&F_{ab}^L=[D_a^L,A_b^L]-[D_b^L,A_a^L]+[A_a^L,A_b^L]-\epsilon_{abc}A_c^L \,,
\\&F_{ab}^R=[D_a^R,A_b^R]-[D_b^R,A_a^R]+[A_a^R,A_b^R]-\epsilon_{abc}A_c^R\,,
\\&F_{ab}^{L,R}=[D_a^L,A_b^R]-[D_b^R,A_a^L]+[A_a^L,A_b^R]\,.
 \end{split} 
 \end{equation}
 Adapting the discussion, starting with equation (\ref{constraint}), it can be seen that only four of these six gauge fields constitute independent degrees of freedom in the commutative limit, $\ell_L,\,\ell_R\rightarrow \infty$.
 
The emerging model has the structure of a $U(n)$ gauge theory on ${\cal M}\times S_F^{2 \, Int}\times S_F^{2\, Int}$ with the gauge fields $A_M=(A_{\mu}\,,A_a)$ and corresponding field strength tensor $F_{MN}=(F_{\mu \nu}\,,F_{\mu a}^L\,,F_{\mu a}^R\,, F_{ab}^L\,,F_{ab}^R\,,F_{ab}^{L,R})$. We can quickly glance over some of the essential features of the $U(4)$ gauge theory on ${\cal M}\times S_F^{2 \, Int}\times S_F^{2\, Int}$. 

For the $U(4)$ theory, taking the symmetry generators $\omega_a^L$ and $\omega_a^R$
\begin{multline}
 \omega_a^L=(X_a^{(2\ell_L+1)}\otimes \bm{1}^{(2\ell_R+1)}\otimes \bm{1}_4\otimes \bm{1}_4)+(\bm{1}^{(2\ell_L+1)}\otimes \bm{1}^{(2\ell_R+1)}\otimes \tilde{\Gamma_a^0}^L\otimes \bm{1}_4)
 \\-(\bm{1}^{(2\ell_L+1)}\otimes \bm{1}^{(2\ell_R+1)}\otimes \bm{1}_4\otimes i\frac{L_a^L}{2})\,,
 \end{multline}
 \begin{multline}
 \omega_a^R=(\bm{1}^{(2\ell_L+1)}\otimes X_a^{(2\ell_R+1)}\otimes \bm{1}_4\otimes \bm{1}_4)+(\bm{1}^{(2\ell_L+1)}\otimes \bm{1}^{(2\ell_R+1)}\otimes \tilde{\Gamma_a^0}^R\otimes \bm{1}_4)
 \\-(\bm{1}^{(2\ell_L+1)}\otimes \bm{1}^{(2\ell_R+1)}\otimes \bm{1}_4\otimes i\frac{L_a^R}{2})\,,
 \end{multline} 
 with $(L_a^L,L_a^R)$ same as before, we can construct the $SU(2)\times SU(2)$-equivariant fields. $SU(2)\times SU(2)$ representation content of $(\omega_a^L,\omega_a^R)$ follows from the Clebsch-Gordan expansion
\begin{align}
  (\ell_L,\ell_R)\otimes\left((\frac{1}{2},0)\oplus(\frac{1}{2},0)\right)\otimes (\frac{1}{2},\frac{1}{2}) &\equiv \bm{2}(\ell_L,\ell_R+\frac{1}{2})\oplus\bm{2}(\ell_L,\ell_R-\frac{1}{2})\oplus \bm{2}(\ell_L+\frac{1}{2},\ell_R)\nonumber
  \\&\oplus \bm{2}(\ell_L-\frac{1}{2},\ell_R)\oplus (\ell+1,\ell_R-\frac{1}{2})\oplus  (\ell+1,\ell_R+\frac{1}{2})\nonumber
\\&\oplus(\ell-1,\ell_R-\frac{1}{2})\oplus(\ell-1,\ell_R+\frac{1}{2})\oplus(\ell_L-\frac{1}{2},\ell_R-1)\nonumber
\\&\oplus (\ell_L+\frac{1}{2},\ell_R-1)\oplus (\ell_L+\frac{1}{2},\ell_R+1)\nonumber
\\&\oplus (\ell_L-\frac{1}{2},\ell_R+1)\nonumber
\\& := I\,.
\label{gaugerep}
\end{align}
$\Pi_{\pm}^L\,,\Pi_{\pm}^R\in Mat((2\ell_L+1)\times(2\ell_R+1)\times4\times4)$ project to the representations in the decomposition (\ref{gaugerep}) as given in the table below.
\begin{center}
    \begin{tabular}{c | c }
    Projector & To the Representation \\ \hline
    \\
    ${\Pi}_{\pm}^L = \frac{1}{2} (\pm i Q^L_I + {\Pi}^L_{\frac{1}{2}})$ & $(\ell_L,\ell_R+\frac{1}{2})\oplus (\ell_L,\ell_R-\frac{1}{2})\oplus(\ell_L\pm 1,\ell_R+\frac{1}{2})\oplus(\ell_L\pm1,\ell_R-\frac{1}{2})$ \\
    ${\Pi}_{\pm}^R = \frac{1}{2} (\pm i Q^R_I + {\Pi}^R_{\frac{1}{2}})$ & $(\ell_L+\frac{1}{2},\ell_R )\oplus (\ell_L+\frac{1}{2},\ell_R\pm 1 )\oplus  (\ell_L-\frac{1}{2},\ell_R )\oplus(\ell_L-\frac{1}{2},\ell_R\pm1 )$
\end{tabular}
\end{center} The $SU(2)\times SU(2)$-equivariance conditions indicate that $A_\mu\,,A_a^L\,,A_b^R$ satisfy the relevant adapted version of (\ref{conditions}). As before, we can determine the dimensions of solution spaces for $A_{\mu},\,\,A^L_a$ and $A^R_a$ using the Clebsch-Gordan decomposition of the adjoint action of $(\omega_a^L,\omega_a^R)$. We find
\begin{align}
 I\otimes I\equiv \bm{24} (0,0)\oplus \bm{52}(1,0)\oplus \bm{52}(0,1)\oplus \cdots.
 \label{I}
\end{align}
This means that there are $24$-invariants. The solution space for each of $A_a^L,\,A_a^R$ is $52$-dimensional. We further see that there are no spinor representations $(\frac{1}{2},0)$ or $(0,\frac{1}{2})$ occuring in (\ref{I}). This corroborates perfectly with our initial expectations, in view of the fact that $({\Gamma_a^0}^L,{\Gamma_a^0}^R)$ cannot be expressed through a bilinear of fields with the desired symmetry properties. If the latter was possible, it would have contradicted the absence of the equivariant spinor field modes and vice versa.

A suitable set of $24$ invariants is given by the following matrices
\begin{align}
 & \Pi^L_+\,,\quad Q^L_+\,,\quad \Pi^L_-\,,\quad Q^L_-\,,\quad \Pi^R_+\,,\quad Q^R_+\,,\quad \Pi^R_-\,,\quad Q^R_-\,,\quad Q_F^L\,,\quad Q_H^L\,,\quad Q_F^R\,,\quad Q_H^R\,,\nonumber
 \\&\Pi^L_+Q_B^R\,,\quad \Pi^L_-Q_B^R\,,\quad \Pi^R_+Q_B^L\,,\quad \Pi^R_-Q_B^L\,,\quad Q_+^LQ_B^R\,,\quad Q_-^LQ_B^R\,,\quad Q_F^LQ_B^R\,,\quad Q_H^LQ_B^R\,,\nonumber
 \\&Q_+^RQ_B^L\,,\quad Q_-^RQ_B^L\,,\quad Q_F^RQ_B^L\,,\quad Q_H^RQ_B^L\,,
 \label{invariants}
 \end{align}
where $Q^L_\pm$, $Q^L_F$, $Q^L_H$, $Q^L_{BI}$ are in same formal form as \eqref{i2} and likewise for the set of matrices $Q^R$. 

A set of $52$ linearly matrices transforming under the $(1,0)$ representation may be provided as
\begin{align}
&[D_a^L,Q_+^L]\,,\quad Q_+^L[D_a^L,Q_+^L]\,,\quad \lbrace D_a^L,Q_+^L\rbrace\,,\quad
Q_B^R[D_a^L,Q_+^L]\,,\quad Q_B^RQ_+^L[D_a^L,Q_+^L]\,,\quad Q_B^R\lbrace D_a^L,Q_+^L\rbrace\,,\nonumber
\\&[D_a^L,Q_-^L]\,,\quad Q_-^L[D_a^L,Q_-^L]\,,\quad \lbrace D_a^L,Q_-^L\rbrace\,,\quad
Q_B^R[D_a^L,Q_-^L]\,,\quad Q_B^RQ_-^L[D_a^L,Q_-^L]\,,\quad Q_B^R\lbrace D_a^L,Q_-^L\rbrace\,,\nonumber
 \\&[D_a^L,Q_F^L]\,,\quad Q_F^L[D_a^L,Q_F^L]\,,\quad \lbrace D_a^L,Q_F^L\rbrace\,,\quad
Q_B^R[D_a^L,Q_F^L]\,,\quad Q_B^RQ_F^L[D_a^L,Q_F^L]\,,\quad Q_B^R\lbrace D_a^L,Q_F^L\rbrace\,,\nonumber
 \\&[D_a^L,Q_H^L]\,,\quad Q_H^L[D_a^L,Q_H^L]\,,\quad \lbrace D_a^L,Q_H^L\rbrace\,,\quad
Q_B^R[D_a^L,Q_H^L]\,,\quad Q_B^RQ_H^L[D_a^L,Q_H^L]\,,\quad Q_B^R\lbrace D_a^L,Q_H^L\rbrace\,,\nonumber
 \\&\Pi_{+}^R[D_a^L,Q_B^L]\,,\quad \Pi_{+}^RQ_B^L[D_a^L,Q_B^L]\,,\quad \Pi_{+}^R\lbrace D_a^L,Q_B^L\rbrace\,,\quad Q_{+}^R[D_a^L,Q_B^L]\,,\quad Q_{+}^RQ_B^L[D_a^L,Q_B^L]\,,\nonumber
 \\&Q_+^R\lbrace D_a^L,Q_B^L\rbrace\,,\quad \Pi_{-}^R[D_a^L,Q_B^L]\,,\quad \Pi_{-}^RQ_B^L[D_a^L,Q_B^L]\,,\quad \Pi_{-}^R\lbrace D_a^L,Q_B^L\rbrace\,,\quad Q_{-}^R[D_a^L,Q_B^L]\,,\nonumber
 \\&Q_{-}^RQ_B^L[D_a^L,Q_B^L]\,,\quad Q_-^R\lbrace D_a^L,Q_B^L\rbrace\,,\quad Q_{F}^R[D_a^L,Q_B^L]\,,\quad Q_{F}^RQ_B^L[D_a^L,Q_B^L]\,,\quad Q_{F}^R\lbrace D_a^L,Q_B^L\rbrace\,,\nonumber
 \\&Q_{H}^R[D_a^L,Q_B^L]\,,\quad Q_{H}^RQ_B^L[D_a^L,Q_B^L]\,,\quad Q_{H}^R\lbrace D_a^L,Q_B^L\rbrace\,,\quad \Pi_{+}^L\omega_a^L\,,\quad \Pi_{-}^L\omega_a^L\,,\quad Q_B^R\Pi_{+}^L\omega_a^L\,,\nonumber
 \\&Q_B^R\Pi_{-}^L\omega_a^L\,,\quad\Pi_{+}^R\omega_a^L\,,\quad \Pi_{-}^R\omega_a^L\,,\quad Q_{+}^R\omega_a^L\,,\quad Q_{-}^R\omega_a^L\,,\quad Q_{F}^R\omega_a^L\,,\quad Q_{H}^R\omega_a^L
\label{52vector}
\end{align} while a linearly independent set transforming as $(0,1)$ is obtained from (\ref{52vector}) by taking $L \leftrightarrow R$.

Monopole sectors exist in this case too and they can be accessed by projecting from $S_F^{2 \, Int}\times S_F^{2\, Int}$. We have, for instance
\begin{align}
 &S_F{^2}^{L\pm}\times S_F{^{2}}^{R\pm}=\bigg (S_F^2(\ell_L)\times S_F^2(\ell_R\pm\frac{1}{2})\bigg)\oplus \bigg( S_F^2(\ell_L\pm \frac{1}{2})\times S_F^2(\ell_R)\bigg )\label{casei},
\\&S_F{^2}^{L,\,2}\times S_F{^{2}}^{R,\,0}=\bigg (S_F^2(\ell_L+\frac{1}{2})\times S_F^2(\ell_R)\bigg)\oplus \bigg (S_F^2(\ell_L-\frac{1}{2})\times S_F^2(\ell_R)\bigg),\label{caseii}
  \\&S_F{^2}^{L,\,0}\times S_F{^{2}}^{R,\ 2}=\bigg (S_F^2(\ell_L)\times S_F^2(\ell_R+\frac{1}{2})\bigg)\oplus \bigg ( S_F^2(\ell_L) \times S_F^2(\ell_R-\frac{1}{2})\bigg),\label{caseiii}
 \end{align}
with the winding numbers $(\pm1,\pm1),\,(2,0),\,(0,2)$, respectively.

We can project to the $(\pm1,\pm1)$ sector using
 \begin{align}
  (1-\Pi_{\mp}^L)(1-\Pi_{\mp}^R).
 \end{align}
 This projection leaves us with $8$ equivariant scalars
\begin{align}
 \Pi_{\pm}^L\,,\quad \Pi_{\pm}^R\,,\quad Q_{\pm}^L\,,\quad Q_{\pm}^R\,,\quad Q_B^R\Pi_{\pm}^L\,,\quad Q_B^RQ_{\pm}^L\,,\quad Q_B^L\Pi_{\pm}^R\,,\quad Q_B^LQ_{\pm}^R\,,
 \label{sca}
\end{align}
and $16$ vectors carrying the $(1,0)$ representation,
\begin{align}
&[D_a^L,Q_\pm^L]\,,\quad Q_\pm^L[D_a^L,Q_\pm^L]\,,\quad \lbrace D_a^L,Q_\pm^L\rbrace\,,\quad 
Q_B^R[D_a^L,Q_\pm^L]\,,\quad Q_B^RQ_\pm^L[D_a^L,Q_\pm^L]\,,\quad Q_B^R\lbrace D_a^L,Q_\pm^L\rbrace\,,\nonumber
\\&\Pi_{\pm}^R[D_a^L,Q_B^L]\,,\quad \Pi_{\pm}^RQ_B^L[D_a^L,Q_B^L]\,,\quad \Pi_{\pm}^R\lbrace D_a^L,Q_B^L\rbrace\,,\quad Q_{\pm}^R[D_a^L,Q_B^L]\,,\quad Q_{\pm}^RQ_B^L[D_a^L,Q_B^L]\,,\nonumber
\\&Q_{\pm}^R\lbrace D_a^L,Q_B^L\rbrace\,,\quad \Pi_{\pm}^L\omega_a^L\,,\quad Q_B^R\Pi_{\pm}^L\omega_a^L\,,\quad \Pi_{\pm}^R\omega_a^L\,,\quad Q_{\pm}^R\omega_a^L\,,
\label{leftvec}
\end{align}
and another $16$ carrying the $(0,1)$ IRR which are obtained from (\ref{leftvec}) by $L\leftrightarrow R$. 

For the winding number sector $(2,0)$ in the equation (\ref{caseii}), we can use the projection operator
 \begin{align}
 (1-\Pi_{+}^R)(1-\Pi_-^R)\,.
 \end{align}
In this case, the relevant part of the Clebsch-Gordan expansion gives the result $\bm{12}(0,0)\oplus\bm{28}(1,0)\oplus \bm{24}(0,1)$. Equivariant scalars may be given as the following subset of those in (\ref{invariants})
\begin{align}
 &\Pi_{+}^L\,,\quad \Pi_{-}^L\,,\quad Q_{+}^L\,,\quad Q_{-}^L\,,\quad Q_F^L\,,\quad Q_H^L\,,\quad Q_B^R\Pi_{+}^L\,,\quad Q_B^R\Pi_{-}^L\,,\quad Q_B^RQ_{+}^L\,,\quad Q_B^RQ_{-}^L\,,\nonumber
 \\&Q_B^RQ_F^L\,\quad Q_B^RQ_H^L.
\end{align}
$28$ vectors which carry the $(1,0)$ IRR can be given as
\begin{align}
&[D_a^L,Q_+^L]\,,\quad Q_+^L[D_a^L,Q_+^L]\,,\quad \lbrace D_a^L,Q_+^L\rbrace\,,\quad
Q_B^R[D_a^L,Q_+^L]\,,\quad Q_B^RQ_+^L[D_a^L,Q_+^L]\,,\quad Q_B^R\lbrace D_a^L,Q_+^L\rbrace\,,\nonumber
\\&[D_a^L,Q_-^L]\,,\quad Q_-^L[D_a^L,Q_-^L]\,,\quad \lbrace D_a^L,Q_-^L\rbrace\,,\quad 
Q_B^R[D_a^L,Q_-^L]\,,\quad Q_B^RQ_-^L[D_a^L,Q_-^L]\,,\quad Q_B^R\lbrace D_a^L,Q_-^L\rbrace\,,\nonumber
\\&[D_a^L,Q_F^L]\,,\quad Q_F^L[D_a^L,Q_F^L]\,,\quad \lbrace D_a^L,Q_F^L\rbrace\,,\quad
Q_B^R[D_a^L,Q_F^L]\,,\quad Q_B^RQ_F^L[D_a^L,Q_F^L]\,,\quad Q_B^R\lbrace D_a^L,Q_F^L\rbrace\,,\nonumber
\\&[D_a^L,Q_H^L]\,,\quad Q_H^L[D_a^L,Q_H^L]\,,\quad \lbrace D_a^L,Q_H^L\rbrace\,,\quad 
Q_B^R[D_a^L,Q_H^L]\,,\quad Q_B^RQ_H^L[D_a^L,Q_H^L]\,,\quad Q_B^R\lbrace D_a^L,Q_H^L\rbrace\,,\nonumber
\\&\Pi_{+}^L\omega_a^L\,,\quad \Pi_{-}^L\omega_a^L\,,\quad Q_B^R\Pi_{+}^L\omega_a^L\,,\quad Q_B^R\Pi_{-}^L\omega_a^L\,.
\end{align}
while there are $24$ matrices which carry the $(0,1)$ IRR and they may be listed as
\begin{align}
 &\Pi_{+}^L[D_a^R,Q_B^R]\,,\quad \Pi_{+}^LQ_B^R[D_a^R,Q_B^R]\,,\quad \Pi_{+}^L\lbrace D_a^R,Q_B^R\rbrace\,,\quad Q_{+}^L[D_a^R,Q_B^R]\,,\quad Q_{+}^LQ_B^R[D_a^R,Q_B^R]\,,\nonumber
 \\ &Q_{+}^L\lbrace D_a^R,Q_B^R\rbrace\,,\quad \Pi_{-}^L[D_a^R,Q_B^R]\,,\quad \Pi_{-}^LQ_B^R[D_a^R,Q_B^R]\,,\quad \Pi_{-}^L\lbrace D_a^R,Q_B^R\rbrace\,,\quad Q_{-}^L[D_a^R,Q_B^R]\,,\nonumber
 \\& Q_{-}^LQ_B^R[D_a^R,Q_B^R]\,,\quad Q_{-}^L\lbrace D_a^R,Q_B^R\rbrace\,,\quad Q_{F}^L[D_a^R,Q_B^R]\,,\quad Q_{F}^LQ_B^R[D_a^R,Q_B^R]\,,\quad Q_{F}^L\lbrace D_a^R,Q_B^R\rbrace\,, \nonumber
 \\&Q_{H}^L[D_a^R,Q_B^R]\,,\quad Q_{H}^LQ_B^R[D_a^R,Q_B^R]\,,\quad Q_{H}^L\lbrace D_a^R,Q_B^R\rbrace\,,\quad \Pi_{+}^L\omega_a^R\,,\quad \Pi_{-}^L\omega_a^R\,,\quad Q_{+}^L\omega_a^R\,,\nonumber
 \\&Q_{-}^L\omega_a^R\,,\quad Q_{F}^L\omega_a^R\,,\quad Q_{H}^L\omega_a^R.
\end{align}
To describe the monopole sectors with the winding number $(0,2)$, it is sufficient to make the exchange $L\leftrightarrow R$.

\end{document}